\documentclass[twocolumn,prb,reprint,showpacs,preprintnumbers,amsmath,amssymb,superscriptaddress]{revtex4-1}
\usepackage{graphicx}
\usepackage{dcolumn}
\usepackage{bm}
\usepackage{graphicx}
\usepackage{subfigure}
\usepackage{physics}
\usepackage{epstopdf}
\usepackage{braket}
\usepackage{placeins}
\usepackage{float}
\usepackage{mathtools}
\usepackage{xcolor}

\newcommand*{\citen}[1]{%
  \begingroup
    \romannumeral-`\x 
    \setcitestyle{numbers}%
    \cite{#1}%
  \endgroup   
}

\begin{document}

\title{Dynamical quantum phase transitions in Weyl semimetals}
\author{Aritra Lahiri}\email{aritra.l895@gmail.com}
\affiliation{School of Physics and Astronomy, University of Minnesota, Minneapolis, MN 55455, USA}
\affiliation{Department of Electrical Engineering, Indian Institute of Technology Bombay, Powai, Mumbai-400076, India\\}
\author{Soumya Bera}\email{soumya.bera@phy.iitb.ac.in}
\affiliation{Department of Physics, Indian Institute of Technology Bombay, Powai, Mumbai-400076, India\\}
\date{\today}

\medskip
\widetext
\begin{abstract}
The quench dynamics in type-I inversion symmetric Weyl semimetals (WSM) are explored in this work which, due to the form of the Hamiltonian, may be readily extended to two-dimensional Chern insulators. We analyze the role of equilibrium topological properties characterized by the Chern number of the pre-quench ground state in dictating the non-equilibrium dynamics of the system, specifically, the emergence of dynamical quantum phase transitions (DQPT). By investigating the ground state fidelity, it is found that a change in the signed Chern number constitutes a sufficient but not necessary condition for the occurrence of DQPTs. Depending on the ratio of the transverse and longitudinal hopping parameters, DQPTs may also be observed for quenches lying entirely within the initial Chern phase. Additionally, we analyze the zeros of the boundary partition function discovering that while the zeros generally form two-dimensional structures resulting in one-dimensional critical times, infinitesimal quenches may lead to one-dimensional zeros with zero-dimensional critical times provided the quench distance scales appropriately with the system size. This is strikingly manifested in the nature of non-analyticies of the dynamical free energy, revealing a logarithmic singularity. In addition, following recent experimental advances in observing the dynamical Fisher zeros of the Loschmidt overlap amplitude through azimuthal Bloch phase vortices by Bloch-state tomography, we rigorously investigate the same in WSMs. Finally, we establish the relationship between the dimension of the critical times and the presence of dynamical vortices, demonstrating that only one-dimensional critical times arising from two-dimensional manifolds of zeros of the boundary partition function lead to dynamical vortices.
\end{abstract}
\maketitle
\section{\label{intro}Introduction}
Non-equilibrium dynamics have garnered significant attention due to experimental advances in cold atomic systems~\cite{Songeaao4748,Leder2016,RevModPhys.83.863,Eisert2015,annurev-conmatphys-031214-014548,RevModPhys.80.885,doi:10.1080/00018732.2010.514702,PhysRevLett.119.080501} and quantum simulators~\cite{Kreula2016,Bernien2017}, permitting the realization of topological phases of matter via engineered synthetic gauge fields~\cite{Goldman6736,Jotzu2014,Goldman2016}. Transcending the equilibrium paradigm, systems driven into non-equilibrium show non-analyticities in dynamical/temporal behavior of observables~\cite{PhysRevLett.110.135704,PhysRevLett.115.140602,PhysRevB.89.125120,PhysRevLett.119.080501,PhysRevLett.113.205701,PhysRevB.93.085416,PhysRevB.96.180303,PhysRevE.81.020101,PhysRevB.97.174303,doi:10.1063/1.4969869,PhysRevLett.113.265702,PhysRevB.89.161105}. This phenomenon has been termed dynamical quantum phase transition (DQPT) (see Refs.~\citen{heylprog} and~\citen{1811.02575} for a review) and it has also recently been experimentally demonstrated in a variety of systems~\cite{PhysRevLett.119.080501,Flaschner2018, 1806.09269,1807.04483, 1806.10871, 1808.03930}. Of particular interest is drawing the connections between non-equilibrium dynamics and equilibrium topology which concerns itself with the classification of non-interacting matter in equilibrium using topological invariants, wherein one aims to correlate the emergence of DQPTs with changes in the topological invariants in non-interacting two-band insulators~\cite{PhysRevLett.115.236403,PhysRevB.94.155104,PhysRevB.91.155127,PhysRevB.95.184307,PhysRevB.93.085416,PhysRevLett.117.086802,PhysRevLett.118.185701}. With the rapid progress in the field of topology in condensed matter physics, topological properties are no longer limited to gapped systems. For instance, Weyl semi-metals (WSM), arising from split Dirac crossings due to the breaking of inversion and/or time-reversal symmetries, have emerged as the quintessential representative of gapless topological systems~\cite{PhysRevX.5.011029,PhysRevX.5.031013,PhysRevB.84.235126,Huang2015,Jia2016}. Three dimensional WSM are topologically protected semi-metallic quantum states harboring the elusive Weyl fermions in condensed matter systems as low-energy excitations at the Weyl nodes. In reciprocal space they can be seen as magnetic monopoles which act as sources or sinks of Berry curvature. At the surfaces of a WSM slab, this results in the emergence of topologically protected chiral states called Fermi arcs, joining the Weyl nodes of opposite chirality in the region where Chern number equals one. Due to the rich Chern phase structure in three dimensions and the consequent additional degree of freedom as well as the manifestation of bulk topology in the surface Fermi arcs, WSMs provide the perfect avenue to study the interplay between equilibrium topology and DQPTs.

In this work, firstly, the bulk effects of quantum quenches are analyzed by studying the role of topology in governing the occurrence of DQPTs. We perform sudden quenches in type-I inversion symmetric single WSMs~\cite{PhysRevB.93.075108,PhysRevB.93.201302}, studying the dynamics using the boundary partition function which, for real times describes the Loschmidt overlap (LO) amplitude as mentioned in Eq.~\eqref{bpf} and the corresponding discussion. It has been shown earlier that for two-dimensional systems described by the Chern number, a change in the magnitude of the Chern number is a necessary condition to observe a DQPT~\cite{PhysRevB.91.155127,PhysRevB.95.184307} while, in Sec~\ref{secdqptcond} we find that a change in the signed Chern number constitutes a sufficient but not necessary condition for the existence of pre-quench state wavevectors which trigger DQPTs. We explain this observation by relating the excitation probability of the effective two level system, which must equal half (infinite temperature) to trigger a DQPT~\cite{PhysRevB.93.144306,PhysRevB.95.184307}, to the ground state fidelity across the topological phase transition. This treatment provides a window in the domain of the quenched parameter where DQPTs occur, while also highlighting the role of crossing Chern phase boundaries or equivalently encountering Weyl nodes along the quench path. Moreover, studies on both integrable~\cite{PhysRevB.89.161105,PhysRevB.89.125120} and non-integrable models~\cite{PhysRevB.92.104306} have shown that the occurrence of DQPTs need not bear a direct correspondence with equilibrium phase transitions. In this regard, we show that in WSMs, DQPTs may also be observed for quenches entirely lying in the initial Chern phase depending on the ratio of the transverse and longitudinal hopping parameters. We, therefore, provide a comprehensive characterization of the quench conditions for observing DQPTs and the connections with the topological quantum critical point. Additionally, in Sec~\ref{seclozero} we analyze the zeros of the boundary partition function which are composed of multiple tails with each tail corresponding to a Chern phase boundary encountered in the quench path. Consequently, the orientation of the tails depend of the quench path as the number of imaginary axis intersections of the zeros forming the critical times for DQPTs must match the number of Chern phase boundaries crossed. The dimension of the tails and their intersection with the imaginary axis (critical times for DQPTs) are found to depend not only on the system dimension, but also on the quench parameters and quench distance. An otherwise two-dimensional tail of zeros originating from the given quench forming a one-dimensional critical time manifold, converts into a one-dimensional tail with a zero-dimensional (point-like) critical time for small quenches near the Chern phase boundares suitable scaled with the system size. Clearly, it follows that the nature of the non-analyticities of the free-energy, which depend on the dimension of the critical times, evolve with the quench protocol. Further, recent advances in experimental techniques permit the real time observation of the dynamical zeros of the LO in particular from the dynamical free energy~\cite{PhysRevLett.119.080501} and using full time- and momentum-space Bloch-state tomography~\cite{Flaschner2018}. Motivated by these, we first study in the dynamical free energy non-analyticities in Sec~\ref{secfreeen}, revealing the nature of non-analyticities at the critical times. Subsequently, in Sec~\ref{secdynvort}, following the experimental observation~\cite{Flaschner2018} of dynamical vortices in the Bloch azimuthal phase of quenched states in the Haldane model using Bloch-state tomography, we study and analyze the same in WSM. We derive expressions for the vorticity of the dynamical vortices and establish the relation between the dimension of critical times and the presence of dynamical vortices, discovering that only one dimensional critical times arising from two-dimensional manifolds of zeros of the boundary partition function lead to dynamical vortices. 

Lastly, the form of the Hamiltonian of type-I inversion symmetric WSMs, as described in the following section, permits a simple extension of the conclusions of this work to two-dimensional Chern insulators in general.

\section{Model}
\label{Model}
\begin{figure}
\begin{center}
\includegraphics[width=0.85\linewidth]{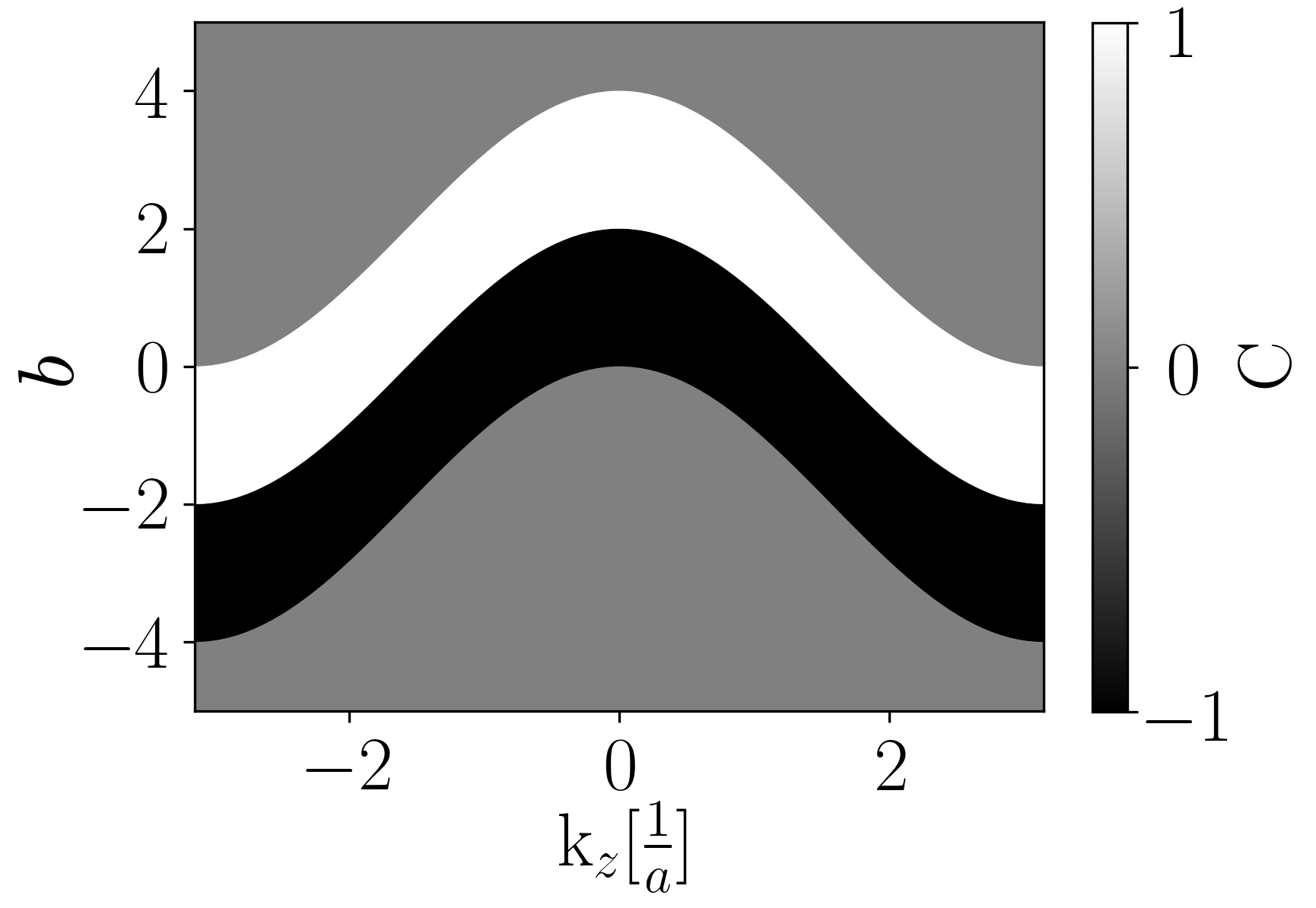}
\caption{Chern number phase diagram for the type-I inversion symmetric single WSM lattice model~\eqref{wsm1}. The equations of the boundaries are obtained from the degeneracies beginning with the uppermost one are $\mathrm{cos}(k_z)=\frac{t'}{2t}(2-b)$, $\mathrm{cos}(k_z)=\frac{t'}{2t}(-b)$ and $\mathrm{cos}(k_z)=\frac{t'}{2t}(-2-b)$.}
\label{Chernphasewsm1}
\end{center}
\end{figure}

We consider the following two-band model for non-interacting fermions with translation invariance of a type-I inversion symmetric single WSM~\cite{PhysRevB.93.075108,PhysRevB.93.201302} described by the Hamiltonian,
\begin{align}
\mathcal{H}&=\frac{1}{2}\sum_{\bm{x};s=1,2}c^\dagger_{\bm{x}+\bm{a}_{s}}\left[it\sigma_s-\frac{t'}{2}\sigma_3\right]c_{\bm{x}}+\frac{bt'}{2}\sum_{\bm{x}}c^\dagger_{\bm{x}}c_{\bm{x}}\nonumber\\
&+\frac{t}{2}\sum_{\bm{x}}c^\dagger_{\bm{x}+\bm{a}_{3}}c_{\bm{x}}+\text{h.c.}. \label{wsm1lattice}
\end{align}
which represents stacked Chern-insulator layers coupled along the $\hat{z}(s=3)-$direction~\cite{PhysRevB.93.075108,PhysRevLett.115.246603}. Here $c,c^\dagger$ denote the fermion operators, $\sigma_j$ are the Pauli matrices, $t(t')$ denote the hopping parameters, $b$ is the control parameter being quenched, and $\bm{a}_s$ are cubic-lattice unit vectors along the direction indexed by $s$. In the momentum-space, this Hamiltonian translates to $\mathcal{H}=\sum_{\bm{k}} H(\bm{k})=\sum_{\bm{k}} \bm{h}(\bm{k})\cdot\bm{\sigma}$ with,
\begin{align}
h_n&=t\mathrm{sin}(k_n) \qquad(n=x,y) \nonumber\\
h_z&=t\mathrm{cos}(k_z)+\frac{t'}{2}\left(b-\mathrm{cos}\left(k_x\right)-\mathrm{cos}\left(k_y\right)\right)\label{wsm1}
\end{align}
Since the system is essentially constituted by layers of Chern insulators, one for each value of $k_z$, the conclusions of this work are applicable to Chern insulators. The ground state of $\mathcal{H}$ is given by the product state of the ground states at each momentum: $|\Psi_{-}\rangle=\prod_{\bm{k}}|g(\bm{k})\rangle$. The topology of the ground state of this Hamiltonian is captured by the first Chern number $C=\frac{1}{2\pi}\iint_{\mathrm{BZ}(k_z)} d\bm{k}\frac{\bm{h}}{2h^3}\left(\partial_{k_x}\bm{h}\times \partial_{k_y}\bm{h}\right)$, as shown in Fig.~\ref{Chernphasewsm1}. The Chern phase boundaries, beginning with the uppermost one (as shown in Fig.~\ref{Chernphasewsm1}) are located at the degeneracies given by $(k_x,k_y)=(0,0),$ $(0,\pm\pi)$ or $(\pm\pi,0)$, and $(\pm\pi,\pm\pi)$, respectively. Accordingly, the boundaries are described by $\mathrm{cos}(k_z)=\frac{2t'}{t}(-b)$ and $\mathrm{cos}(k_z)=\frac{2t'}{t}(\pm 2- b)$, respectively.

We study the non-equilibrium real-time evolution and DQPTs following a sudden quench of the parameter $b$ as described in~\eqref{wsm1}, from $b_i$ to $b_f$ and thereby changing the Hamiltonian from $H_i(\bm{k})=\bm{h_i}(\bm{k})\cdot\bm{\sigma}$ to $H_f(\bm{k})=\bm{h_f}(\bm{k})\cdot\bm{\sigma}$. Due to the translational invariance, different momentums are decoupled and are analyzed separately. At half filling, the system is prepared in the ground state (lower-Bloch band) $\lvert\Psi_i(\bm{k})\rangle=\lvert g(\bm{k},b_i)\rangle$ of the initial Hamiltonian $H_i(\bm{k})$. During the course of the quench, the eigen-basis changes with the system parameters determining the quench. For a sudden quench beginning with the initial ground state, the post-quench state is expressed as $|\Psi_i(\bm{k})\rangle=\lvert g(\bm{k},b_i)\rangle=u_{\bm{k}}\lvert g(\bm{k},b_f)\rangle + v_{\bm{k}}\lvert e(\bm{k},b_f)\rangle$, which equals the pre-quench ground state ($\lvert\Psi_f(\bm{k})\rangle=\lvert\Psi_i(\bm{k})\rangle$) with a change of basis. We characterize the dynamics of the system using the boundary partition function with the boundary state being the pre-qench ground state, which is expressed as~\cite{PhysRevLett.110.135704,PhysRevB.93.144306,PhysRevB.95.184307,PhysRevB.89.161105,PhysRevB.89.125120,PhysRevLett.101.120603}
\begin{align}
G(z)&=\langle \Psi_i |e^{-z\mathcal{H}_f}|\Psi_i\rangle=\prod_{\bm{k}}\langle\Psi_i(\bm{k}) |e^{-zH_f(\bm{k})}|\Psi_i(\bm{k})\rangle
\end{align}
Now, the boundary partition function for each momentum is given by,
\begin{align}
G(z,\bm{k})&=\langle\Psi_i(\bm{k})|e^{-zH_f(\bm{k})}\lvert\Psi_i(\bm{k})\rangle\nonumber\\
&=|u_{\bm{k}}|^2e^{+zh_f(\bm{k})}+|v_{\bm{k}}|^2e^{-zh_f(\bm{k})}\nonumber\\
&=\left( (1-p_{\bm{k}})+p_{\bm{k}}e^{-2zh_f(\bm{k})} \right)e^{zh_f(\bm{k})} \label{bpf}
\end{align} 
where $p_{\bm{k}}=|v_{\bm{k}}|^2=\frac{1}{2}\left(1-\frac{\bm{h_i}(\bm{k})}{h_i(\bm{k})}\cdot\frac{\bm{h_f}(\bm{k})}{h_f(\bm{k})}\right)$ denotes the probability of excitation following the quench, $|u_{\bm{k}}|^2=1-|v_{\bm{k}}|^2$, and $+(-)h_f(\bm{k})$ is the energy of $|e(\bm{k},b_f)\rangle\big(|g(\bm{k},b_f)\rangle\big)$ under $H_f(\bm{k})$. Note that $p_{\bm{k}}$ depends on the quench protocol. For imaginary $z=it$, the Loschmidt overlap (LO)~\cite{PhysRevA.30.1610,PhysRevLett.110.135704,PhysRevB.93.144306,PhysRevB.95.184307,PhysRevB.89.161105,PhysRevB.89.125120,PhysRevLett.101.120603} amplitude is obtained from Eq.~\eqref{bpf} as $G(it,\bm{k})=\mathrm{cos}(th_f)+i\left(\frac{\bm{h}_i\cdot\bm{h}_f}{h_ih_f}\right)\mathrm{sin}(th_f)$. Thus, the real time zeros of the boundary partition function or equivalently the zeros of the LO are given by $t_n=(2n+1)\frac{\pi}{2E_f(\bm{k}^*)}$, where $p_{\bm{k}^*}=\frac{1}{2}$.

Analogous to the thermal free energy, the dynamical partition function scales exponentially with the system volume. Hence, one may define a dynamical free energy for non-interacting fermions as, 
\begin{align}
f(z)&=-\displaystyle{\lim_{L\rightarrow\infty}}\frac{1}{L^d}\mathrm{ln}(G(z))\nonumber\\&=-\int \frac{d\bm{k}}{(2\pi)^3}\mathrm{ln}\left( (1-p_{\bm{k}})+p_{\bm{k}}e^{-2zh_f(\bm{k})} \right) \label{dqptf}
\end{align}
where the $L^d$ is the system volume. DQPTs are therefore contingent upon the existence of purely imaginary (real time) non-analyticities of the dynamical free energy~\cite{PhysRevLett.110.135704,PhysRevLett.115.140602,PhysRevB.89.125120,PhysRevLett.119.080501}, corresponding to the zeros at critical times of the LO. Now, the non-analyticities of the free energy, or the zeros of the LO, are given by
\begin{align}
z_n=\frac{1}{2h_f(\bm{k})}\left[\mathrm{ln}\left(\frac{p_{\bm{k}}}{1-p_{\bm{k}}}\right)+i\pi(2n+1)\right] \label{lozero}
\end{align} 
which are obtained when~\cite{PhysRevB.93.144306,PhysRevB.95.184307} there exist $\bm{k}=\bm{k}^*$ such that $p_{\bm{k}^*}=1/2$. This condition defines the sectors in momentum-space ($\bm{k}$) where DQPTs may occur. Note that the exact calculation of $p_{\bm{k}}$ depends on the quenching protocol.

\section{Conditions for DQPT}
\label{secdqptcond}
\begin{figure}[h!]
\begin{center}
\includegraphics[width=0.99\linewidth]{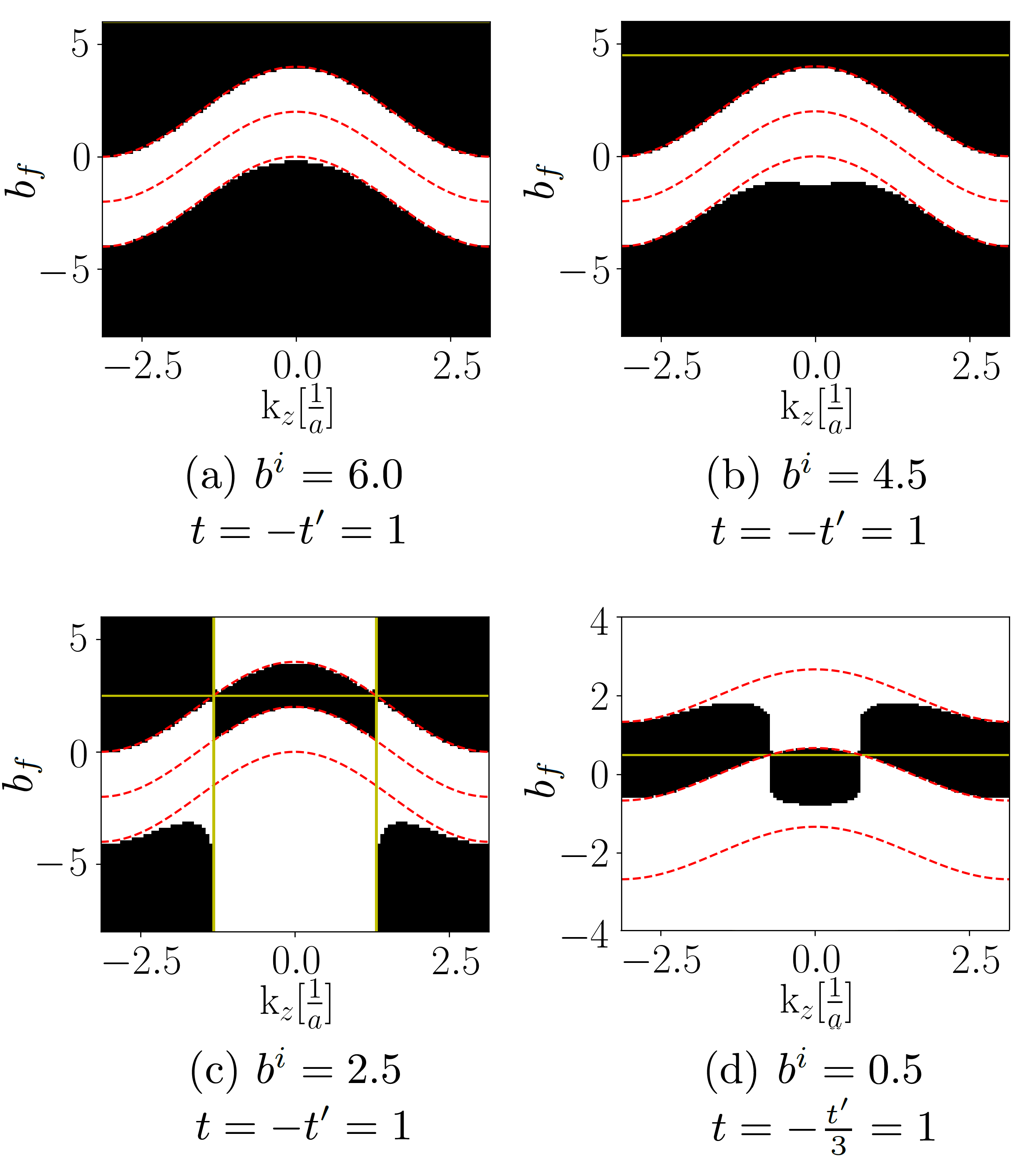}
\caption{The domains in parameter space for the Hamiltonian~\eqref{wsm1} with $t=-t'=1$ which yield DQPTs for pre-quench (a) $b_i=6.0$, (b) $b_i=4.5$ and (c) $b_i=2.5$. Regions shown in white denote the occurrence of a DQPT. Red lines show the Chern phase boundaries (Fig.~\ref{Chernphasewsm1}), while the yellow line shows the $b_i$. A change in the Chern number guarantees a DQPT. Note that quenches from the upper $C=0$ phase to the lower $C=0$ phase in general doesn't trigger a DQPT except in a small region near the $C=0/-1$ phase boundary. This region is marked by the green vertical lines. (d) Same as before, but for $t=-\frac{t'}{3}=1$ which shows DQPTs even for quenches lying entirely within the initial Chern phase.}
\label{SWSMdqpt}
\end{center}
\end{figure}

We calculate the excitation probabilities $p_{\bm{k}}$ for the WSM model~\eqref{wsm1} for the quench $b=(b_i\to b_f)$ over the entire two-dimensional Brillouin zone for different values of $k_z$ to determine the sectors yielding DQPTs. A quench performed at a given value of $k_z=k_{z_{0}}$ is deemed as yielding a DQPT if there exists atleast one wavevector $\bm{k}(k_{z_{0}})=(k_x,k_y,k_{z_{0}})$ in the corresponding 2D Brillouin zone yielding a DQPT, i.e., $p_{\bm{k}(k_{z_{0}})}=\frac{1}{2}$. Fig.~\ref{SWSMdqpt} shows these sectors as a function of $k_z$ and $b_f$ for different values of $b_i$, revealing the role of equilibrium topology. We present four quenches which capture the physics adequately. It is clear from all the panels of Fig.~\ref{SWSMdqpt} that DQPTs necessarily occur when the quench is accompanied by a change in the signed Chern number. Equivalently, Weyl nodes  have to be crossed along the quench path to observe DQPTs as the phase boundary between each pair of Chern phases is associated with the appearance of Weyl nodes. For instance, in Fig.~\ref{SWSMdqpt}(c) where $b_i=2.5$ lies in the $C=1$ phase at $k_z=0$, DQPTs can be observed only when the post-quench $b=b_f$ lies in a phase with $C\neq 1$, i.e., $b_f\in (-\infty, 2)\cup(4, \infty)$ corresponding to the Chern phase boundaries described by $\mathrm{cos}(k_z)=\frac{t'}{2t}(2-b)$ and $\mathrm{cos}(k_z)=-\frac{t'}{2t}(b)$ with $t=-t'=1$. Note that these phase boundaries correspond to the degeneracies of the Hamiltonian and are therefore tantamount to the appearance of Weyl nodes. 
Further, as seen from all the panels, quenches from the upper(lower) $C=0$ phase to the lower(upper) $C=0$ phase generally don't lead to DQPTs unless $b_i$ and $b_f$ both lie sufficiently close to the phase boundaries bordering the pre and post-quench Chern phases. In Fig.~\ref{SWSMdqpt}(a), where $b_i$ lies deep within the upper $C=0$ phase, on quenching to the lower $C=0$ phase DQPTs occur only in an extremely small region close to the phase boundary. This region enlarges when $b_i$ decreases and moves closer to the $C=0(\text{upper})/C=1$ phase boundary, as seen in Fig.~\ref{SWSMdqpt}. Note that this region yielding DQPTs exists only for those values of $k_z$ for which $b_i$ is sufficiently close to the $C=0(\text{upper})/C=1$ phase boundary, which in this case occurs for $k_z$ close to zero.

Further, the inversion symmetry of the system, as evident from the Hamiltonian given by~\eqref{wsm1}, gives us a relation between the wavevectors yielding DQPTs. Consequently,
\begin{align}
h(\bm{k})&=h(-\bm{k})\label{invymprop2}\\
\bm{h_i}(\bm{k_\parallel},k_z)\cdot \bm{h_f}(\bm{k_\parallel},k_z)&=\bm{h_i}(-\bm{k_\parallel},k_z)\cdot \bm{h_f}(-\bm{k_\parallel},k_z)\label{invsymprop2}
\end{align}
where $\bm{k_\parallel}=(k_x,k_y)$. These ensure that the if the state with wavevector $(k_x,k_y,k_z)$ yields a DQPT at some time, then concurrently the state with wavevector $(-k_x,-k_y,k_z)$ also yields a DQPT.

\subsection{Ground-state fidelity analysis}
The conditions mentioned above are less restrictive than the one derived by Vajna and D\'ora~\cite{PhysRevB.91.155127} namely, a change in the absolute value of the Chern number is a necessary condition to observe a DQPT. In order to provide an intuitive explanation for the conditions found above for sudden quenches, we turn to the ground state fidelity~\cite{PhysRevB.81.012303,gritpolkovch}. The fidelity, $F(\lambda,\lambda+d\lambda)=\langle\Psi(\lambda+d\lambda)\lvert\Psi(\lambda)\rangle$, is the overlap between the ground states of infinitesimally parametrically perturbed Hamiltonians. Clearly, $F(\lambda,\lambda)=1$. 
\begin{figure}[H]
\begin{center}
\includegraphics[width=\linewidth]{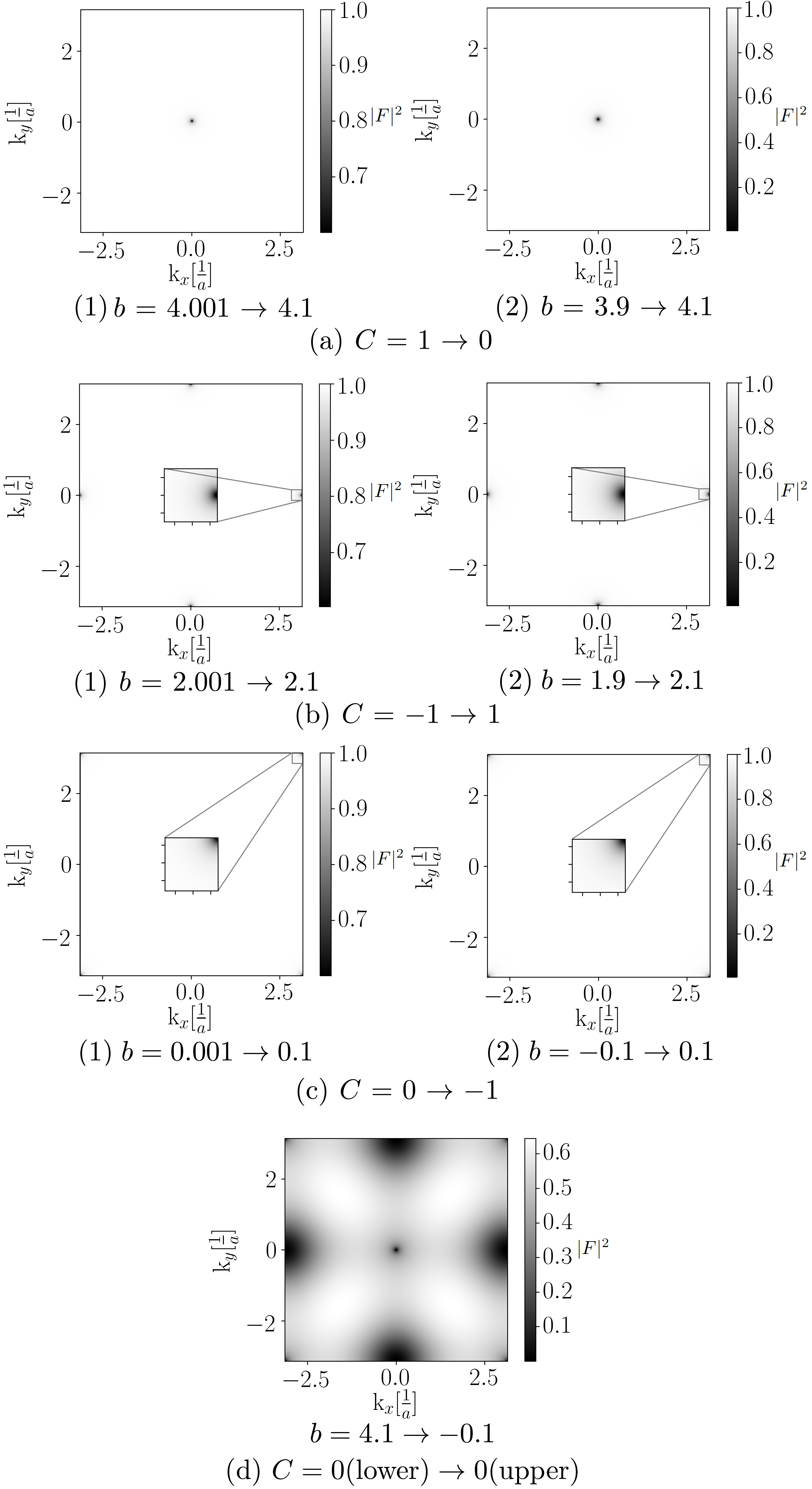}
\caption{$|F|^2$ for the quenches described in the captions, at $k_z=0$. In each sub-figure, the panels ($1$) shows quenches with the same initial and final Chern phases while the panels ($2$) shows Chern phase boundary crossing. Only in panels ($2$), at the $(k_x,k_y)$ values sufficiently close to the corresponding Chern phase boundary degeneracies, $|F|^2$ crosses half and triggers DQPTs. In (d), for the quench from the upper to the lower Chern zero phases, $(k_x,k_y)$ corresponding to all the boundaries lying in the quench path, which in fact includes all the Chern boundaries, contribute.}
\label{SWSMdqptFsq}
\end{center}
\end{figure}
Being a measure of the distinguishability of the infinitesimally perturbed states, it exhibits a sharp drop in its value (which is otherwise close to one) at the critical point, thereby characterizing quantum phase transitions~\cite{1751-8121-41-41-412001,PhysRevLett.99.100603,PhysRevLett.99.095701,PhysRevE.76.022101,PhysRevLett.96.140604,gritpolkovch}. The fidelity may now be written as $|F(\lambda,\lambda+d\lambda)|^2=1-|\langle\Psi_g(\lambda+d\lambda)\lvert\Psi_g(\lambda)\rangle|^2=1-p_{\bm{k}}$, implying that the occurrence of a DQPT following the quench $\lambda\to\lambda+d\lambda$ is synonymous with $|F(\lambda,\lambda+d\lambda)|^2$ dropping below $1/2$. By expanding the fidelity for infinitesimal quenches and focusing on the leading term, one obtains $|F(\lambda,\lambda+d\lambda)|^2=1-\frac{1}{2}g_{\mu\nu}d\lambda^\mu d\lambda^\nu=1-\frac{1}{2}(\gamma_{\mu\nu}-\beta_\mu\beta_\nu)d\lambda^\mu d\lambda^\nu$, where $\gamma_{\mu\nu}=\langle\partial_\mu\Psi(\lambda)|\partial_\nu\Psi(\lambda)\rangle$ is the real part of the quantum geometric tensor and $\beta_{\mu}=\langle\Psi(\lambda)|\partial_\mu\Psi(\lambda)\rangle$ is the Berry connection~\cite{provost1980}. Using first order perturbation theory one obtains, 
\begin{align}
g_{\mu\nu}(\bm{k})&=\sum_{n\neq0}\mathrm{Re}\frac{\langle\Psi_0(\lambda)|\partial_\mu H\lvert\Psi_n(\lambda)\rangle\langle\Psi_n(\lambda)|\partial_\nu H\lvert\Psi_0(\lambda)\rangle}{(E_0 - E_n)^2}\nonumber\\
&=\frac{|\langle\Psi_g(\lambda)|\partial_\mu H\lvert\Psi_e(\lambda)\rangle|^2}{4h^2}=\frac{t'^2(h^2-h_z^2)}{16h^4} \label{gmetricwsm}
\end{align} 
where the second equality employs the particle-hole symmetry of the problem and the third equality requires the explicit use of the Hamiltonian given by~\eqref{wsm1} with $\lambda=b$. Therefore, for an infinitesimal quench $\delta \lambda=\delta b\to 0$ at a particular $\bm{k}$, $|F(b,b+\delta b, \bm{k})|^2=1-\frac{t'^2(h^2-h_z^2)}{32h^4}(\delta b)^2=1-\chi_F(\bm{k})(\delta b)^2$. Our objective will be to show that for the quench $b_i\to b_f$, $|F|^2$ necessarily goes below 0.5 on crossing a phase boundary. Equivalently, the fidelity susceptibility must 

The fidelity susceptibility $\chi_F=\frac{g}{2}$ for infinitesimal quenches across a Chern phase transition is divergent at the phase boundary, as shown below. This leads to a significant contribution to the fidelity squared when quenching across the Chern phase boundary. Using the location of the Chern phase boundaries in the 2D-BZ as described in Sec.~\ref{Model} for the boundary between $C=1$ and $C=0$ phases,
\begin{align}
\chi_F(\bm{k})&=\frac{t'^2t^2\left(k_x^2+k_y^2\right)}{32\left[t^2\left(k_x^2+k_y^2\right)+\frac{t'^2}{4}\left((b-b_0)+\frac{k_x^2}{2}+\frac{k_y^2}{2}\right)^2\right]^2}\label{chiexp00}
\end{align}
For $\bm{k}=\left(\kappa\mathrm{cos}(\gamma), \kappa\mathrm{sin}(\gamma), \mathrm{cos}^{-1}\left(\frac{t'}{2t}(2-b_0)\right)\right)$ with $b_0=4$,
\begin{align}
\chi_F(\kappa)&=\frac{t'^2t^2\kappa^2}{32\left[t^2\kappa^2+\frac{t'^2}{4}\left((b-b_0)+\frac{\kappa^2}{2}\right)^2\right]^2}\label{chiexp0}\\
&\xrightarrow{\kappa\to 0}\frac{\pi^2 t^2}{8\left(t^2+\frac{t'^2}{4}(b-b_0)\right)}(\delta(b-b_0))^2\label{chiexp}
\end{align}
where $b_0$ is the location of the degenaracy at the given value of $k_z$. Similarly, near the remaining Chern phase boundaries, $\chi_F(\kappa)$ is described by the same expression as in~\eqref{chiexp}. When summed over the entire 2D-BZ at a given value of $k_z$, $|F(b,b+\delta b, k_z)|^2=\prod_{\bm{k}\in \mathrm{2D-BZ}}\left(1-\frac{t'^2(h^2-h_z^2)}{32h^4}(\delta b)^2\right)=1-\chi_F(k_z)(\delta b)^2$ with $\chi_F(k_z)=\sum_{\bm{k}\in \mathrm{2D-BZ}}\chi_F(\bm{k})=\sum_{\bm{k}\in \mathrm{2D-BZ}}\frac{t'^2(h^2-h_z^2)}{32h^4}$. Since $\chi_F(\bm{k})$ is significant only near the degeneracies where~\eqref{chiexp0} is a good approximation, and rapidly vanishes away from them, we may extend the integration limits (summation replaced by integration in thermodynamic limit) to $\pm\infty$ to obtain,
\begin{align}
\frac{\chi_F(k_z)(2\pi)^2}{(L^d)^2}&\approx \iint \limits_{-\infty}^{+\infty} d^2\bm{k_\parallel}\chi_F(\bm{k})\approx \frac{\frac{\pi t^2}{t'^2}}{(b-b_0)^2}+\mathcal{O}\left(\frac{1}{|b-b_0|}\right)\label{chiexpintg}
\end{align}
where $L$ is the length of the $d-$dimensional system. The dominant contribution is $\chi_F(k_z)\sim\frac{1}{(b-b_0)^2}$ when $b$ is close to the quantum critical point $b_0$. From~\eqref{chiexpintg}, it is once again clear that as expected, the fidelity susceptibility is singular at the Chern phase boundary.

Now, with the Hamiltonian described by $\bm{h}=h(\mathrm{sin}(\theta)\mathrm{cos}(\phi), \mathrm{sin}(\theta)\mathrm{sin}(\phi),\mathrm{cos}(\theta))$, the ground state is given by $|\Psi\rangle=\begin{bmatrix}
\mathrm{sin}\left(\frac{\theta}{2}\right)e^{-i\phi},&-\mathrm{cos}\left(\frac{\theta}{2}\right) 
\end{bmatrix}^T$. For our choice of Hamiltonian, only $h_z$ and therefore only $\theta$ changes. Hence, the general expression for the fidelity is given by,
\begin{align}
F(\theta_i, \phi;\theta_f,\phi)&=\langle\Psi(\theta_f,\phi)\lvert\Psi(\theta_i,\phi)\rangle=\mathrm{cos}\left(\frac{\theta_f-\theta_i}{2}\right)\label{Fsqfull0}\\
|F(\theta_i, \phi;\theta_f,\phi)|^2&=\frac{1}{2}+\frac{h_i^2+(\Delta h_z) h_{z,i}}{2h_fh_i} \label{Fsqfull}
\end{align}
where the subscripts $i(f)$ denote the pre(post)-quench quantities and $\Delta h_z = \frac{t'(b_f-b_i)}{2}$ refers to the change in $h_z$ due to the quench.

For any quench which crosses a phase boundary, the quench path can be decomposed into three parts: the first part traverses the initial phase and reaches infinitesimally close to a phase boundary; second part crosses the boundary and moves infinitesimally into the next phase; and the third part continues thereafter into the final phase. As we shall see in the following analysis, depending on the relative magnitude of the hopping parameters $t$ and $t'$ DQPTs may be triggered either by the first part or the second part. Therefore, a quench crossing a Chern phase boundary constitutes a sufficient but not a necessary condition for the occurrence of DQPTs.

First, we look at the general quench $b_i\to b_f$. Any $h_z \in (-\infty, \infty)$, is bijectively mapped to  $\theta\in(\pi,0)$. From~\eqref{Fsqfull0}, the occurrence of DQPT is contingent upon the condition $|F|^2=\mathrm{cos}^2\left(\frac{\theta_i-\theta_f}{2}\right)=\frac{1}{2}$ or equivalently $\lvert\theta_i-\theta_f\rvert=\frac{\pi}{4}\hspace*{0.1cm}\mathrm{or}\hspace*{0.1cm}\frac{3\pi}{4}$. This defines a window $\lvert\theta_i-\theta_f\rvert = \frac{3\pi}{4}$ beyond which DQPTs can't occur, providing an absolute bound on the largest permissible quench. Clearly, a sudden quench from $b_i=\infty$ to $b_f=-\infty$ can't trigger a DQPT as the corresponding fidelity squared $|F|^2=\mathrm{cos}^2\left(\frac{\pi}{2}\right)=0\hspace*{0.1cm}\forall\hspace*{0.1cm}(k_x,k_y)$, never crossing $\frac{1}{2}$ at any $(k_x,k_y)$. This corroborates the inferences made earlier from Fig.~\ref{SWSMdqpt} regarding quenches from the upper $C=0$ phase to the lower $C=0$ phase, where we noted that $b_i$ and $b_f$ must be sufficiently close to the phase boundaries of the corresponding phases. Now, the system is quenched from $b=b_i$ to $b=b_f=b_i-\Delta b$. First, we look at the phase boundary between $C=1$ and $C=0$ occurring at $\bm{k}=(\kappa, \kappa, k_z)\rvert_{\kappa\to 0}$ with $b_0=2-\frac{2t}{t'}\mathrm{cos}(k_z)$ (see Sec.~\ref{Model}). From Eq.~\ref{Fsqfull}, on performing a Taylor expansion for small values of $\kappa$, the occurrence of DQPT ($|F|^2$ crossing half) requires,
\begin{align}
\Delta b&=\underbrace{\left[b_i-\left(2-\frac{2t}{t'}\mathrm{cos}(k_z)\right)\right]}_{\Delta b_0}\nonumber\\&  + \underbrace{\left[\frac{8t^2}{t'^2\left(b_i-\left(2-\frac{2t}{t'}\mathrm{cos}(k_z)\right)\right)}+1\right]}_{\Delta b_1}\kappa^2+ \mathcal{O}(\kappa^4). \label{epsilonkappa}
\end{align}
Similar expressions can be derived for the remaining Chern phase boundaries. For instance, at the Chern phase boundary between the $C=1$ and the $C=-1$ phases, which occurs at $(k_x,k_y)=(0,\pm\pi)$ or $(k_x,k_y)=(\pm\pi,0)$, one obtains, 
\begin{align}
\Delta b&=\underbrace{\left[b_i-\left(-\frac{2t}{t'}\mathrm{cos}(k_z)\right)\right]}_{\Delta b_0}\nonumber\\&  + \underbrace{\left[\frac{8t^2}{t'^2\left(b_i-\left(-\frac{2t}{t'}\mathrm{cos}(k_z)\right)\right)}-1\right]}_{\Delta b_1}\kappa^2 + \mathcal{O}(\kappa^4), \label{epsilonkappa2}
\end{align}
while at the Chern phase boundary between $C=-1$ and $C=0$(lower) phases, which occurs at $(k_x,k_y)=(\pm\pi,\pm\pi)$, one obtains,
\begin{align}
\Delta b&=\underbrace{\left[b_i-\left(-2-\frac{2t}{t'}\mathrm{cos}(k_z)\right)\right]}_{\Delta b_0}\nonumber\\&  + \underbrace{\left[\frac{8t^2}{t'^2\left(b_i-\left(-2-\frac{2t}{t'}\mathrm{cos}(k_z)\right)\right)}-1\right]}_{\Delta b_1}\kappa^2+ \mathcal{O}(\kappa^4). \label{epsilonkappa1}
\end{align}
Clearly, $\Delta b$ depends on the ratio of the hopping parameters $t$ and $t'$. Note that, in all these equations, the constant term $\Delta b_0$ represents the distance between the pre-quench $b=b_i$ and the location of the Chern phase boundary corresponding to the degeneracy being investigated.
\subsection{Initial Chern phase left}
\label{cpnl0}
Without loss of generality, here we consider quenches from the upper $C=0$ phase to the $C=1$ phase. In~\eqref{epsilonkappa}, the sign of the term $\Delta b_1$ is given by, $\mathrm{sign}(\Delta b_1)=\mathrm{sign}\left(\frac{|t'|^2}{2}\left(b_i-b_0\right)\right)$ which is positive as $b_i$ lies above the Chern phase boundary located at $b=b_0$. Hence, from~\eqref{epsilonkappa}, we obtain $b_fb_i-\Delta b=b_i-\Delta b_0-\Delta b_1\kappa^2+\ldots\approx b_0-\Delta b_1\kappa^2<b_0$, as $\Delta b_1>0,\forall t,t'$. Therefore, the DQPTs occur only for quenches crossing the Chern phase boundary. Further, from~\eqref{epsilonkappa} it is seen that as $\kappa$, which is a measure of the region $\{\bm{k}:|F|^2\leq 1/2\}$ decreases, the minimum quench distance ($\Delta b$) required to trigger DQPTs along the quench path decreases. In other words, proximity of the pre-quench state to the Chern phase boundary aids the occurrence of DQPTs. The analysis presented above is valid for transverse wavevectors $(k_x,k_y)$ lying close to the Chern phase boundary degeneracy under consideration with the value of $b_0$ at the Chern phase boundary determined using $\mathrm{cos}(k_z)=\frac{t'}{2t}(2-b)$ for any given $k_z$. Fig.~\ref{SWSMdqptFsq} shows $|F|^2=1-p_{\bm{k}}$ for the four distinct kinds of quenches possible. As derived earlier, $|F|^2$ crosses the value half near the wave-vectors satisfying the corresponding Chern phase boundary degeneracies. Consequently, for any given quench crossing a phase boundary, one can define a small region in momentum space described by $\{\bm{k}:|F|^2\leq 1/2\}$ centered around the wavevectors representing the corresponding degeneracies. Therefore, with the pre-quench state lying sufficiently close to a Chern phase degeneracy, DQPTs are triggered on crossing the Chern phase boundaries. Since the Chern phase boundaries are obtained at mutually exclusive values of $(k_x,k_y)$, any initial configuration described by $\{\bm{k},b\}$ may encounter only one degeneracy in its quench path. This result should hold for any Weyl type degeneracy with a linear dispersion. 
\subsection{Initial Chern phase NOT left}
\label{cpnl}
As seen earlier, DQPTs can't be triggered by quenches lying entirely within the initial Chern phase for quenches beginning in the upper $C=0$ phase. However, this is indeed possible for quenches beginning in the $C=-1$, as shown below. 

From~\eqref{epsilonkappa1}, we obtain $b_f=b_i-\Delta b_0-\Delta b_1 \kappa^2+\ldots\approx b_0-\Delta b_1 \kappa^2$. Unlike the case studied in Sec.~\ref{cpnl0}, $\Delta b_1$ may be rendered negative with a suitable choice of hopping parameters $(t, t')$ to obtain $b_f>b_0$. Hence, for these values of hopping parameters the post-quench state lies in the pre-quench Chern phase. Consider a quench from $b_i=-\frac{2t}{t'}\mathrm{cos}(k_z)-\delta$, where $\delta\in(0,2)$ describes the distance of the pre-quench state from the Chern phase boundary between the $C=1$ and $C=-1$ phases located at $b=-\frac{2t}{t'}\mathrm{cos}(k_z)$. As seen from Fig.~\ref{Chernphasewsm1}, this pre-quench state resides in the $C=-1$ phase spanning $b\in\left(-\frac{2t}{t'}\mathrm{cos}(k_z),-2-\frac{2t}{t'}\mathrm{cos}(k_z)\right)$. From~\eqref{epsilonkappa1}, we obtain,
\begin{align}
&\left({\Delta b_1}\right)_{C=-1\to C=0(\mathrm{lower})}<0\nonumber\\
&\implies t'\in \left(-\infty, -\sqrt{\frac{8}{2-\delta}}\right)\cup \left(\sqrt{\frac{8}{2-\delta}}, \infty\right) \label{samephdqptc1}
\end{align}
Note that this is possible only if $\delta<2$, which is expected as $\delta\geq 2$ would imply that the pre-quench state lies outside the initial $C=-1$ phase. Further, the result is independent of the value of $k_z$, provided that the pre-quench $b_i$ is specified in terms of the $k_z-$dependent location of the Chern phase boundary, namely, $b_i=-\frac{2t}{t'}\mathrm{cos}(k_z)-\delta$. 

\begin{figure}[h]
\begin{center}
\includegraphics[width=\linewidth]{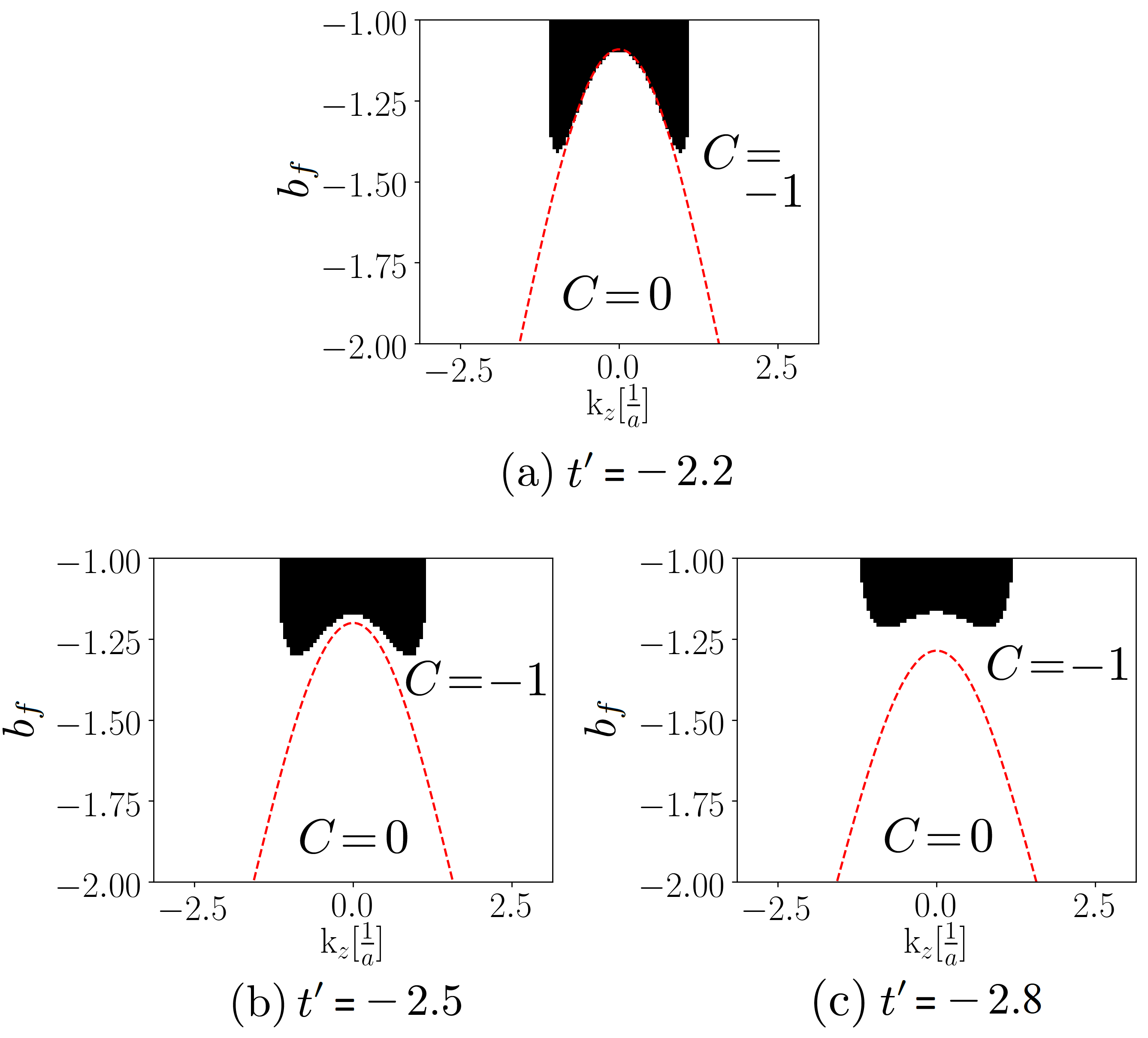}
\caption{The domain in parameter space yielding DQPTs (white), near the Chern phase boundary between the $C=-1$ and $C=0$(lower) phases, for a system described by $t=1$, and (a)$t'=-2.2$, (b) $t'=-2.5$ and (c) $t'=-2.8$. The quench is described by $b_i=-\frac{2t}{t'}-\delta$ with $\delta=0.5$, which equals (a) $0.4091$, (b) $0.3$, and (b) $0.2143$, respectively, for the two panels. With decreasing $t'$, the region yielding DQPTs within the initial $C=-1$ phase enlarges. }
\label{samephdqpt}
\end{center}
\end{figure}

Now, from~\eqref{samephdqptc1}, for $\delta=0.5$ we require $t'\in\left(-\infty, -\sqrt{\frac{8}{2-0.5}}=-2.3094\right)\cup\left(\sqrt{\frac{8}{2-0.5}}=2.3094, \infty\right)$. The triggering of DQPTs for post-quench states lying within the initial Chern phase is illustrated in Figs.~\ref{SWSMdqpt} and~\ref{samephdqpt}. Consequently, at $k_z=0$, we see that DQPTs are observed for post-quench states lying in the $C=-1$ phase for $t'<-2.3094$. Note that the location of the Chern phase boundary depends on $k_z$. As a result, in order to extend the analysis performed here for $k_z=0$ to other values of $k_z$ while keeping the value of $\delta$ same, $b_i$ must be altered accordingly. Similar behavior, where DQPTs may not bear a direct correspondence with equilibrium phase transitions, has previously been noted only for a few systems such as the integrable XY model~\cite{PhysRevB.89.161105}, XXZ model~\cite{PhysRevB.89.125120}, and the non-integrable transverse-field Ising model~\cite{PhysRevB.92.104306}. 

\subsection{Summary}
To summarize, a DQPT occurs when the quench path leaves the pre-quench Chern phase with a pre-quench $(k_x,k_y)$ located sufficiently close to a Chern boundary degeneracy lying along the quench path. Therefore, in the entire 2D Brillouin zone, only the pre-quench $(k_x,k_y)$ located sufficiently close $(0,0); (0,\pm\pi), (\pm\pi,0); (\pm\pi,\pm\pi)$, which represent the Chern boundaries, actually trigger a DQPT on quenching across them. Further, any quench path may cross only one degeneracy as the boundaries are located at mutually exclusive values of $(k_x,k_y)$. Hence, DQPTs occur not only on crossing the boundary between neighboring Chern phases but also on going over multiple Chern phases, for which there must exist $(k_x,k_y)$ in the pre-quench phase such that one of the degeneracies/Chern boundaries is encountered in the quench path. This is observed while quenching from the upper $C=0$ phase to the lower $C=0$ phase as this quench path may cross the degeneracy at the $C=0\to1$ or $C=1\to-1$ or the $C=-1\to1$ boundary depending on the pre-quench $(k_x,k_y)$. Moreover, depending on the ratio of the hopping parameters $t$ and $t'$, DQPTs may even be observed for quenches lying entirely within the initial Chern phase while still being sufficiently close to a Chern phase boundary. 

\section{Zeros of LO}
\label{seclozero}
\begin{figure}[h]
\begin{center}
\includegraphics[width=\linewidth]{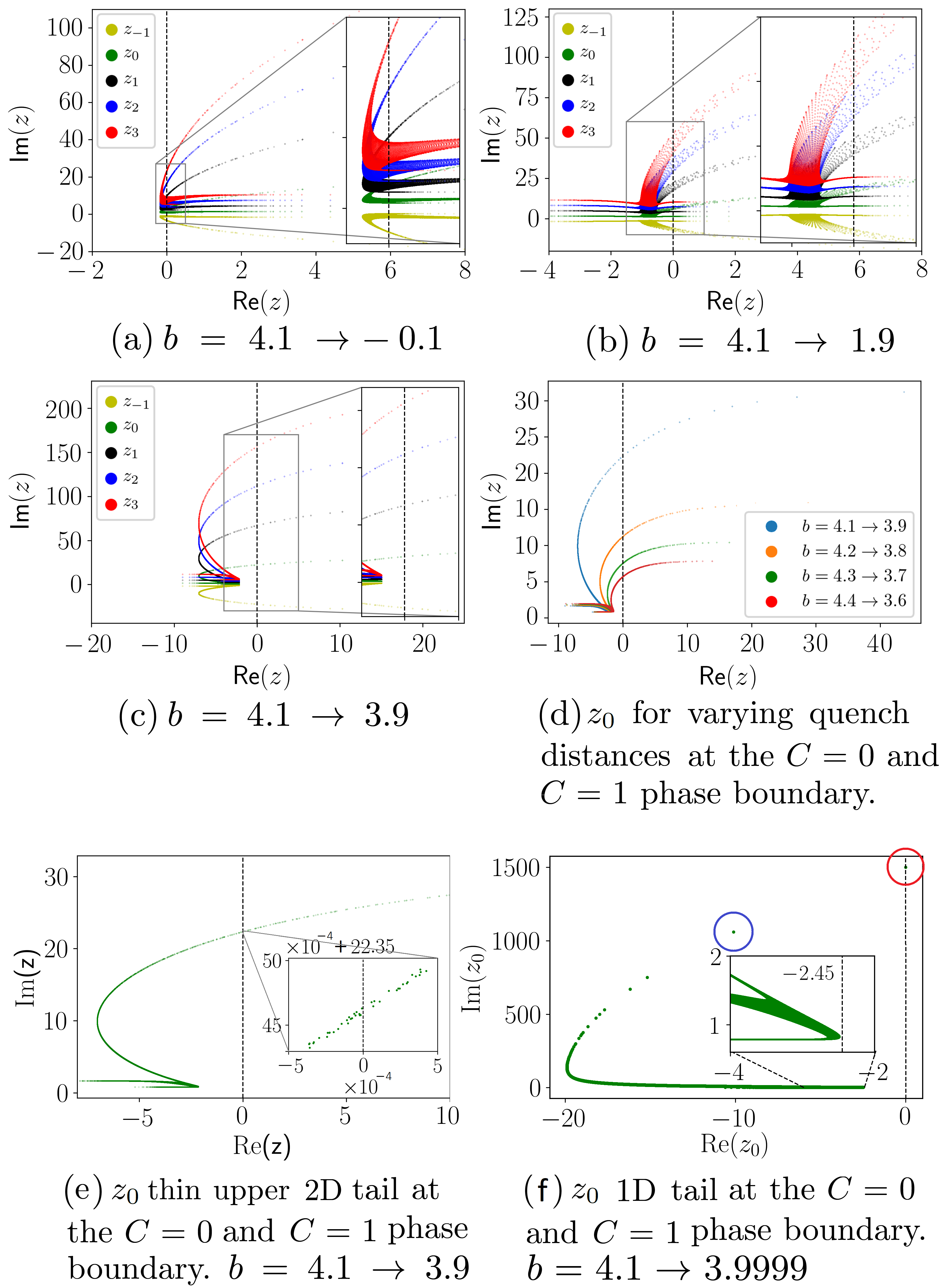}
\caption{System has $t=-t'=1$. (a)-(c) Each set of zeros has three tails, with each tail corresponding to a Chern phase boundary: thin lower tail$-C=0(\mathrm{upper})/1$, thick middle tail$-C=1/-1$ and thin upper tail$-C=-1/0(\mathrm{lower})$. The tails corresponding to the Chern phase boundaries not crossed in the quench move towards the left extending upto $-\infty$, thereby preventing real-time zeros. (d) On increasing the quench distance across the Chern phase boundary (here $C=0/1$ boundary), the wavevectors $\bm{k}^*:p_{\bm{k}^*}=\frac{1}{2}$ move farther away from the degeneracies and thus the real-time intersections occur at smaller times. (e) Zeros of the boundary partition function for the quench $b=4.1\to3.9$ at the $C=0$ and $C=1$ phase boundary with the zoomed inset showing the true 2D nature of the seemingly 1D tail which leads to 1D critical times. (f) 0D critical time (marked in red) in $z_0$ for the quench described by $b=4.1\to 4.1-\Delta b=3.9999$ with $\Delta b$ given by~\eqref{deltab0d} in a system of size $N=6001$. Dashed line in inset shows the zero on the lower tail closest to the imaginary axis with $\text{Re }z_0=-2.45$. }
\label{swsmzero}
\end{center}
\end{figure}

The zeros of the LO, as seen earlier from the discussion following~\eqref{bpf}, correspond to the real times when DQPTs or equivalently non-analyticities in the dynamical free energy are observed. Hence, an analysis of the critical times naturally necessitates an investigation of the zeros of the boundary partition function. In the following sections we proceed to show the structure of the zeros and the dimension of the critical times. The key results that we proceed to show are that the structure and imaginary axis intersection of the zeros correspond to the Chern phase boundaries crossed and dimension of the zeros depend not only on the system dimension, but also on the chosen quench protocol. The zeros are given by $z_n=\frac{1}{2E_f(\bm{k})}\left[\mathrm{ln}\left(\frac{p_{\bm{k}}}{1-p_{\bm{k}}}\right)+i\pi(2n+1)\right]$ are shown in Fig.~\ref{swsmzero} for a sudden quenches with $k_z=0$ and $t=-t'=1$. 

\subsection{Structure of zeros}
\begin{figure}[h]
\begin{center}
\includegraphics[width=.85\linewidth]{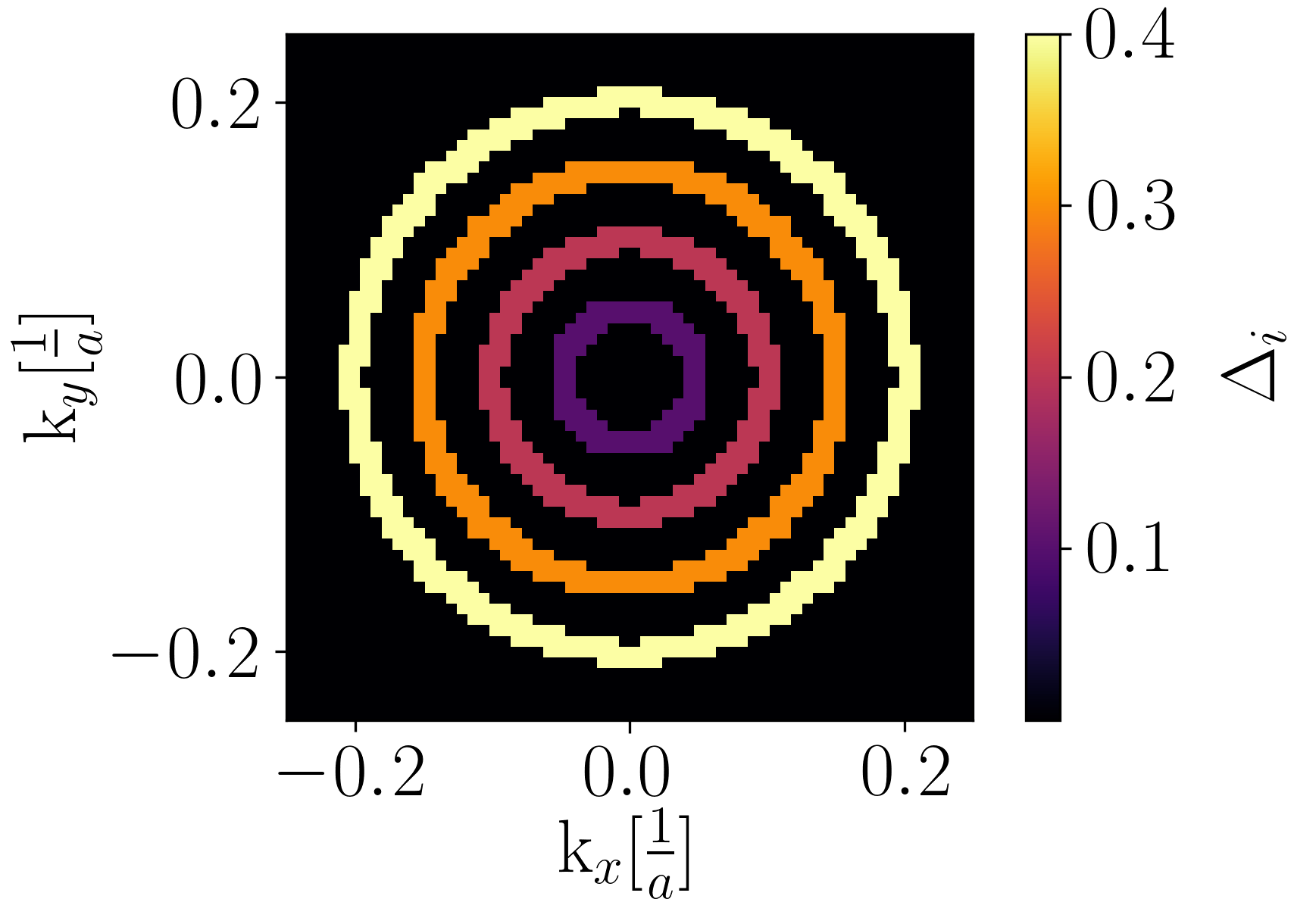}
\caption{Iso-circles of $\left\{(k_x,k_y):p_{\bm{k}}=\frac{1}{2}\right\}$ at the $C=0/1$ Chern phase boundary. The concentric circles centered at $(k_x,k_y)=(0,0)$, which is the point of degeneracy, with increasing radius and lighter colors correspond to $\Delta_i=0.1, 0.2, 0.3$ and $0.4$ respectively. Note that $\kappa=\frac{\Delta_i}{2\sqrt{2}}$ corresponds to $(k_x,k_y)=(\kappa,\kappa)$, implying that the corresponding iso-circle has a radius $\kappa\sqrt{2}=\frac{\Delta_i}{2}$ as seen from the figure.}
\label{kappaDeltai}
\end{center}
\end{figure}

Each set of zeros has a three pronged structure, with each disjoint set of values of $\left\{\bm{k}^*:p_{\bm{k}^*}=\frac{1}{2}\right\}$ leading to an intersection of the corresponding prong of each set of zeros with the imaginary axis. Therefore, the number of imaginary axis intersections is equal to the number of degeneracies/Chern phase boundaries encountered in the quench path. For the quenches considered in Fig.~\ref{swsmzero} we observe two kinds of prongs/tails: thin 2D (seemingly 1D) tails forming nearly point-like 1D intersections with the imaginary axis and thick 2D tails forming 1D intersections. Since the critical times are given by $t_n=(2n+1)\frac{\pi}{2E_f(\bm{k}^*)}$, 1D intersections will be formed iff the iso-circles with $\left\{(k_x,k_y)^*:p_{\bm{k}^*}=\frac{1}{2}\right\}$ do not align with the constant energy contours along them. Note that the constant energy contours of the Hamiltonian are dependent on $b$. From Fig.~\ref{SWSMdqptFsq}(a)-(c), we see that $\left\{\bm{k}^*:p_{\bm{k}^*}=\frac{1}{2}\right\}$ are found only very close to the points of degeneracies, implying that, for quenches into the neighboring Chern phases the zeros will be found very close to the degeneracies. Hence, the critical times for observing the non-analyticities will be huge $\left(t_n=(2n+1)\frac{\pi}{2E_f(\bm{k}^*)}\right)$. Also, on moving within a Chern phase from one boundary to other (by varying $b$), the shape of the constant energy contours change as they are dominated by the location of the degeneracies closest to them with respect to the quench parameter $b$. Hence, on increasing the quench distance across a boundary, the wavevectors $\left\{\bm{k}^*:p_{\bm{k}^*}=\frac{1}{2}\right\}$ move farther away from the corresponding degeneracies, as seen from the relation~\eqref{epsilonkappa}. The consequent increase in $E(\bm{k}^*)$ reduces the critical times, as seen from the decreasing intersections of the $z_0$ manifold with the imaginary axis in Fig.~\ref{swsmzero}(d). To analyze the critical times, we use ~\eqref{epsilonkappa}. For a system described by $t=-t'=1$ at the boundary between the $C=0$ and $C=1$ phases, we perform a quench $b_i=b_0-\Delta\to b_i + \Delta$, implying that $\epsilon=2\Delta$. Substituting these in~\eqref{epsilonkappa}, we get $\kappa^2 =-4+\sqrt{16+\Delta^2}\implies \kappa \approx \frac{\Delta}{2\sqrt{2}}$. The variation in $\kappa$ with $\Delta_i$ can be read off from Fig.~\ref{kappaDeltai}. Thus, the energy from the post-quench Hamiltonian is found as, $E_f\approx\frac{\Delta}{2}\sqrt{t^2+t'^2}=\frac{\Delta}{\sqrt{2}}$, implying $t_0=\frac{\pi}{2E_f}=\frac{\pi}{\Delta\sqrt{2}}$. Consequently, for $\Delta=0.1,0.2,0.3$ and $0.4$, we get $t_0\approx 22.2, 11.1, 7.4$ and $5.5$, which matches with Fig.~\ref{swsmzero}(d). The inverse dependence of the critical times on $\Delta_i$ suggests the use of a larger quench distance to excite higher energy quasiparticle modes to get smaller and observable critical times.

Also, for quenches crossing multiple phase boundaries, such as the quench from the upper to the lower Chern zero phase as seen in Fig.~\ref{SWSMdqptFsq}(d), we find $\left\{\bm{k}^*:p_{\bm{k}^*}=\frac{1}{2}\right\}$ located well away from the degeneracies. Therefore, the critical times will be reasonably small. Further, the thickness of the tails decrease with decreasing quench distance across a Chern phase boundary as the iso-circles overlap with different constant energy contours with decreasingly varying energies. This aspect, which may lead to zero-dimensional critical times, is explored in detail in the following section.

\subsection{Zero-dimensional critical times} 
\label{0dtc}
Previously 0D critical times were obtained in a 2D system~\cite{PhysRevB.92.075114}, the Kitaev honeycomb model, by letting the spin-spin coupling along any one direction to vanish to reduce it to an effective 1D system. Alternatively, without physically lowering the system dimension, 0D critical times arising from imaginary-axis intersections of 1D zero tails may be obtained if the locus of critical wavevectors $\left\{\bm{k}^*:p_{\bm{k}^*}=\frac{1}{2}\right\}$ align perfectly with the constant energy $h_f(\bm{k})$ contours as $t_c\propto \frac{1}{h_f(\bm{k}^*)}$. This doesn't occur except for very small quenches across a Chern phase boundary. We perform such a quench across a Chern phase boundary degeneracy described by the transverse wavevector pair $(k_x,k_y)\rvert_{\mathrm{degeneracy}}=\Gamma$ at any given value of $k_z$ by suddenly varying $b_i\to b_i-\Delta b$, where $b=b_0$ gives the Chern phase boundary for the chosen value of $k_z$. Now, for a system of size $N$ along all directions with lattice constant $a$, the separation between the wavevectors along the $j^{th}$ direction are given by $\Delta k_j=\frac{2\pi}{Na}$. The best alignment of $\bm{k}^*$ with the contours of $h_f$ may be achieved when $\bm{k}^*$ form the wavevectors immediately neighboring $\Gamma$ as a consequence of the discrete separation of the wavevectors. On imposing this condition after Taylor expanding~\eqref{Fsqfull}, similar to~\eqref{epsilonkappa}, at $k_z=0$ and $\bm{k}_\parallel'=(k_x,k_y)=\Gamma+\frac{2\pi}{N}(\pm1,0)$ or $=\Gamma+\frac{2\pi}{N}(0,\pm1)$, we obtain the quench distance,
\begin{align}
\Delta b&=\frac{2h_f^2}{t'h_{f,3}}\bigg\rvert_{\stackrel{(k_x,k_y)=\bm{k}_\parallel}{b_i\to b_i-\Delta b}}\label{deltab0dexact}\\
&\approx\left[b_i-\left(2-\frac{2t}{t'}\right)\right] + \left[\frac{4t^2}{t'^2\left(b_i-\left(2-\frac{2t}{t'}\right)\right)}+1\right]\left(\frac{2\pi}{N}\right)^2 \label{deltab0d}
\end{align} 
which decreases rapidly with increasing system size. This relation governs the scaling of the quench-distance with increasing system size to eventually obtain a 0D real critical time from the 1D zero tails. 

\begin{figure}[h]
\begin{center}
\includegraphics[width=\linewidth]{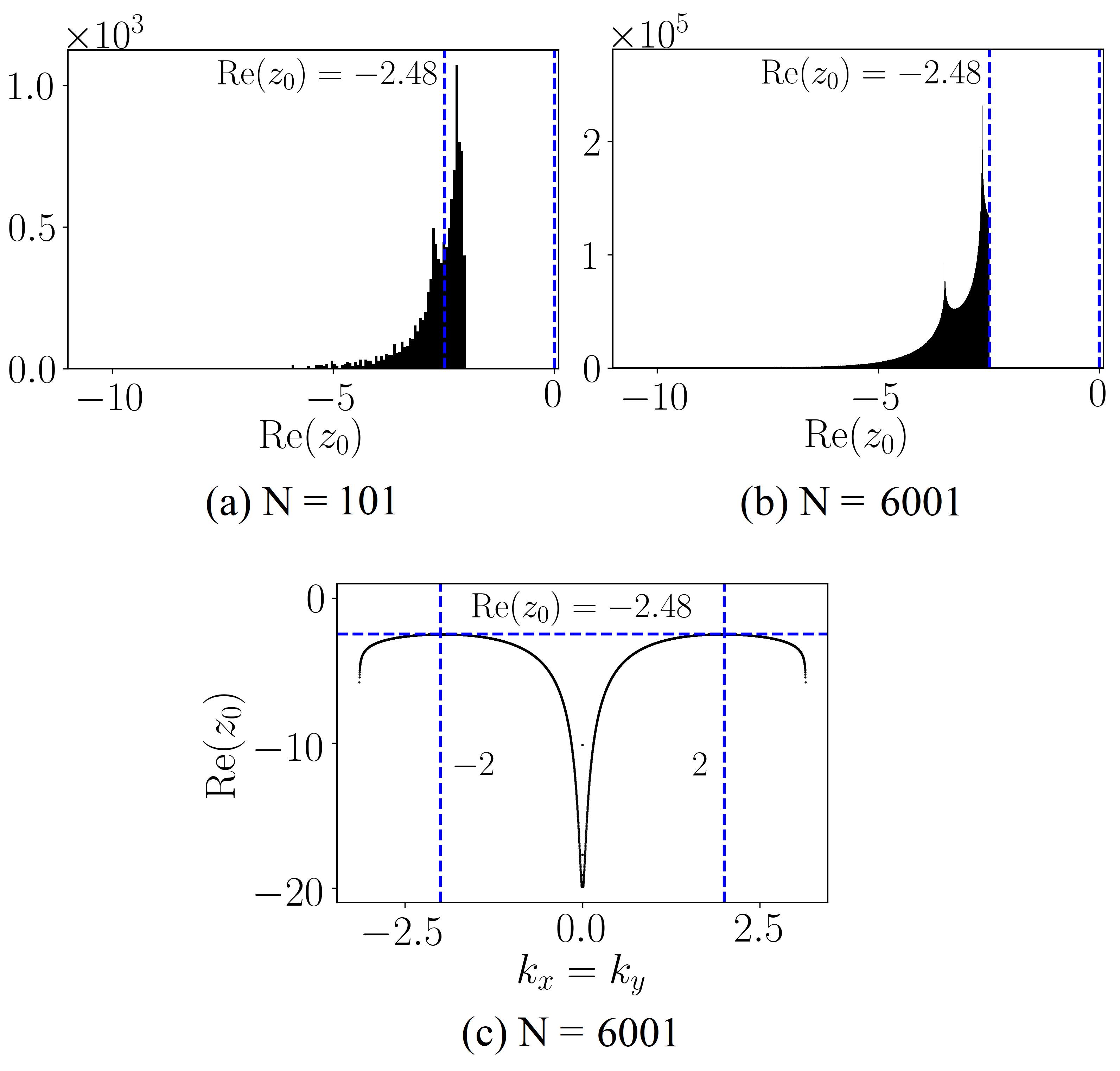}
\caption{Histogram plots of $\mathrm{Re}(z_0)$ evaluated at all the in-plane wavevectors $\bm{k}_\parallel\in$ 2D BZ with $k_z=0$ in a system described by $t=-t'=1$ with system size (a)$N=101$ and (b)$N=6001$, for the quench beginning at $b_1=4.1$ and $b_2=b_1-\Delta b$ as described by~\eqref{deltab0d}. The 0D critical time with $\mathrm{Re}(z_0)=0$ is well isolated from the remaining zeros in the thermodynamic limit, the nearest one of which are located at $\text{Re }z_0\approx-2.45$ (blue dashed line). (c) The real parts of the zeros for $k_xa=k_ya\in [-\pi,\pi]$, showing the wavevector corresponding to the zero immediately neighboring the 0D critical time on the lower tail (See accompanying text and Fig.~\ref{swsmzero}(f) for the structure of the zeros).}
\label{0Dtczeromap}
\end{center}
\end{figure}

The true 0D critical time corresponding to the Chern phase boundary between the $C=0$ and $C=1$ phases at $b_0=4$ is shown in Fig.~\ref{swsmzero}(f) for a quench described by $b_i=4.1$ and $\Delta b$ governed by~\eqref{deltab0d}, performed on a system with $t=-t'=1$ and $N=6001$. Since the excitation probability is directly correlated with the band gap, the imposition of $p_{\bm{k}_\parallel'}=\frac{1}{2}$ at $\bm{k}_\parallel'$ immediately neighboring the degeneracy ensures $p_{\bm{k}}(\bm{k}\in \text{2D-BZ}\backslash\{\Gamma,\bm{k}_\parallel'\})\leq\frac{1}{2}$. Consequently, $\text{Re }z=\mathrm{ln}\left(\frac{p_{\bm{k}}}{1-p_{\bm{k}}}\right)\leq 0$ and therefore, the 0D critical time forms the last finite zero on the sparse 1D tail of zeros. Recall from the preceding discussion that the 0D critical time arising from the imaginary axis intersection of the 1D tail of zeros occurs at $\bm{k}_\parallel'=(k_x,k_y)=\Gamma+\frac{2\pi}{Na}(\pm1,0)$ or $=\Gamma+\frac{2\pi}{N}(0,\pm1)$. As a result, this zero is four-fold degenerate as there are four equivalent wavevectors in the 2D Brillioun zone. 

Further, this 0D critical time is isolated. This may be shown by investigating the location of the zeros immediately neighboring 0D critical time on its left, which come from the two tails (see Fig.~\ref{swsmzero}). 

First, the immediately neighboring zero on the upper tail (marked in blue in Fig.~\ref{swsmzero}) comes from the wavevectors forming the next-nearest neighbors of $(k_x,k_y)=\Gamma$, namely, $\tilde{\bm{k}}_\parallel=(k_x,k_y)=\Gamma+\frac{2\pi}{N}(\pm1,\pm1)$. This identification is made using the knowledge that the zero originating from $\tilde{\bm{k}}_\parallel$ has the second largest finite $\mathrm{Im}(z_0)=\frac{1}{2h_f}$. At these wavevectors, $h_x\approx t\frac{2\pi}{N}$, $h_y\approx t\frac{2\pi}{N}$, $h_{z,i}\approx\frac{t'}{2}(b_i-b_0+\left(\frac{2\pi}{N}\right)^2)$, and $h_{z,f}\approx \frac{t'}{2}\left(b_f-b_0+\left(\frac{2\pi}{N}\right)^2\right)\sim \frac{1}{N^2}$, where we have used~\eqref{deltab0d} to obtain $b_0-b_f\sim\frac{1}{N^2}$. Accordingly, we get, 
\begin{align}
h_i&=\sqrt{h_x^2+h_y^2 +\left(\frac{t'}{2}\left(b_i-b_0+\left(\frac{2\pi}{N}\right)^2\right)\right)^2}\label{hi0dtc}\\
h_f&=\sqrt{h_x^2+h_y^2 +\left(\frac{t'}{2}\Big(b_f-b_0+\Big(\frac{2\pi}{N}\right)^2\Big)\Big)^2}\nonumber\\
&\approx\frac{2\pi t\sqrt{2}}{N}\coloneqq \frac{\Omega}{N}\label{hf0dtc}
\end{align} 
As a result, on using ~\eqref{hi0dtc} and~\eqref{hf0dtc} along with $\Delta h_z$ as obtained from~\eqref{deltab0dexact} in~\eqref{Fsqfull}, 
\begin{align}
p_{\tilde{\bm{k}}}&=1-|F|^2=\frac{1}{2}-\frac{h_i^2+\Delta h_z h_{z,i}}{2h_ih_f}\nonumber\\
&\approx\frac{1}{2}-\underbrace{\left[\frac{\sqrt{2}\pi \left(t^2+\left(\frac{t'}{2}\right)^2\frac{(b_i-b_0)}{2}\right)}{t\lvert t'\rvert(b_i-b_0)}\right]}_{\zeta}\left(\frac{1}{N}\right)+\mathcal{O}(\frac{1}{N^2})\label{pk0dtc}
\end{align} 
where $\zeta=\frac{\sqrt{2}\pi \left(t^2+\left(\frac{t'}{2}\right)^2\frac{(b_1-b_0)}{2}\right)}{t\lvert t'\rvert(b_i-b_0)}$ is a positive constant depending on the system and quench parameters. Hence, from~\eqref{pk0dtc} and~\eqref{hf0dtc},
\begin{align}
\lim_{N\to\infty}\text{Re }z_{\tilde{\bm{k_\parallel}}}&=\lim_{N\to\infty}\frac{1}{2h_f(\tilde{\bm{k_\parallel}})}\mathrm{ln}\left(\frac{p_{\tilde{\bm{k_\parallel}}}}{1-p_{\tilde{\bm{k_\parallel}}}}\right)\nonumber\\
&\approx\lim_{N\to\infty}\frac{1}{2\frac{\Omega}{N}}\mathrm{ln}\left(\frac{\frac{1}{2}-\frac{\zeta}{N}}{\frac{1}{2}+\frac{\zeta}{N}}\right)=-\frac{2\zeta}{\Omega}\nonumber\\
&=-\frac{\left(1+\left(\frac{t'}{2t}\right)^2\frac{(b_1-b_0)}{2}\right)}{\lvert t'\rvert(b_i-b_0)}\label{nzerolim0}
\end{align} 
thereby establishing the sparsity of the upper tail of zeros as $\lim_{N\to\infty}\text{Re }z_{\tilde{\bm{k_\parallel}}}$ is well separated from $0$. Second, for the zeros residing on the lower tail, in the thermodynamic limit $(N\to\infty)$, the real part of the zero immediately neighboring the 0D critical time gravitates to a constant.
For instance, for the case shown in Fig.~\ref{0Dtczeromap}, the real part of the neighboring zero, which comes from the lower tail, is approximately $-2.45$. 

In short, it is concluded that for any finite non-zero $t,t'$, 
\begin{align}
\lim_{N\to\infty}\sup_{\bm{k}\in \text{2D-BZ}\backslash\{\Gamma,\bm{k}_\parallel'\}}\text{Re }z(\bm{k})\neq 0 \label{nzerolim}
\end{align}

In the subsequent section, it is shown that the 0D nature of the critical time is indeed revealed by the behavior of the dynamical free energy as seen in Fig.~\ref{swsmfreeE}(c), which presents a logarithmic singularity concurrent with the isolated nature of the zero. Additionally, the 0D critical time has a distinctive feature of remaining undetected by dynamical azimuthal phase vortices, as described in Sec.~\ref{secdynvort}.

To summarize, an isolated 0D critical time may be obtained by tuning the quench protocol even in a 2D non-interacting system.

\section{Dynamical free-energy non-analyticities}
\label{secfreeen}
\begin{figure}
\begin{center}
\includegraphics[width=\linewidth]{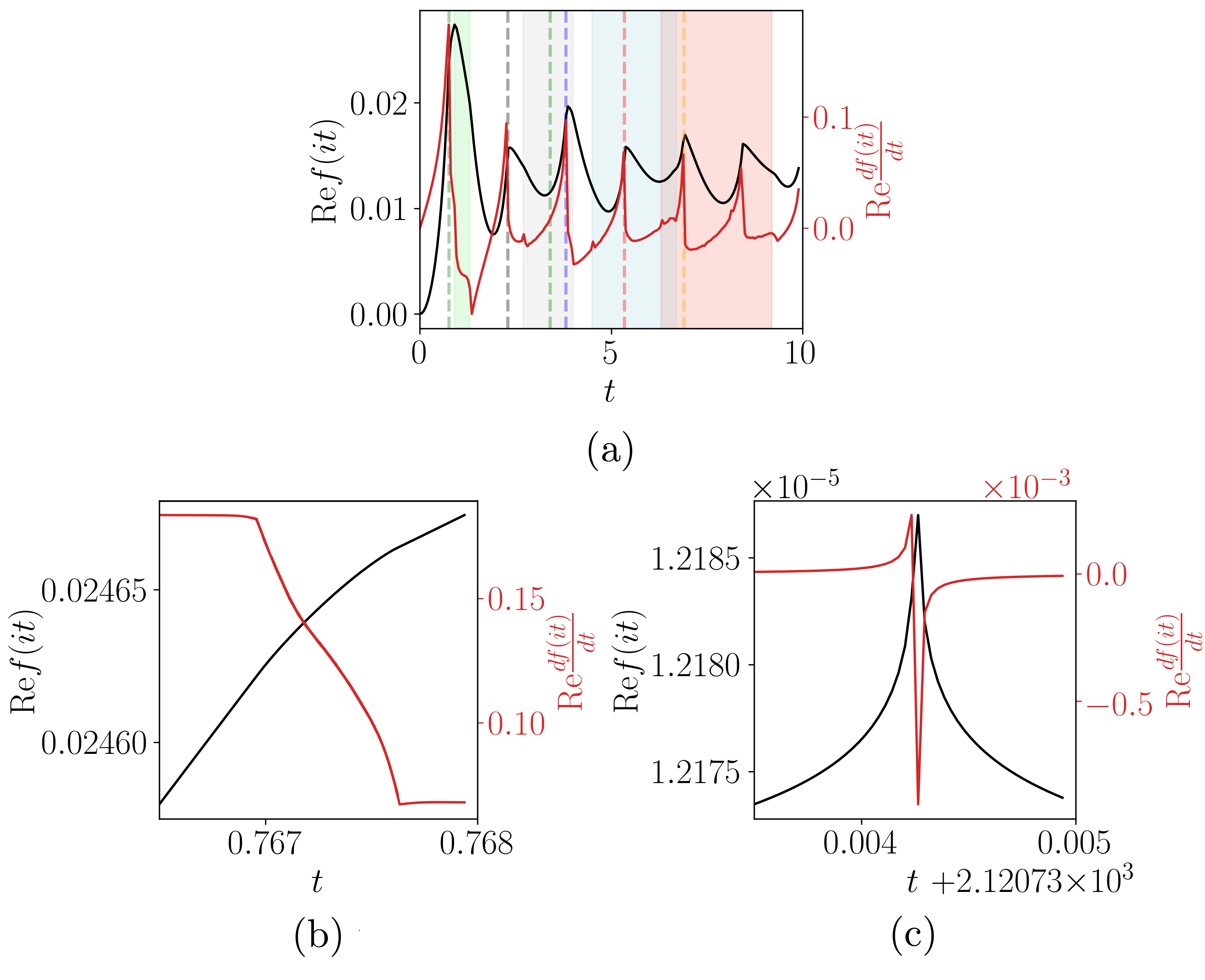}
\caption{(a) $\mathrm{Re}f(z=it)$ (black line) and $\mathrm{Re}f'(z=it)$ (red line) for the same quench as in Fig.~\ref{swsmzero}(a). The colors of the shaded regions and the dashed lines match the colors of the zeros they depict in Fig.~\ref{swsmzero}(a). $\mathrm{Re}f(z=it)$ shows non-analyticities only at the imaginary axis intersections of the zeros shown in Fig.~\ref{swsmzero}(a). (b) $\mathrm{Re}f(z=it)$ (black line) and $\mathrm{Re}f'(z=it)$ (red line) at the seemingly 1D tail at $t\approx 0.767$, denoted by the dashed green line in Fig.~\ref{swsmzero}(a). There are two closely spaced sharp non-differentiable points in $\mathrm{Re}f'(z=it)$, bringing out the true 1D nature of the critical time, establishing that the thin tails are in fact 2D tails. (c) $\mathrm{Re}f(z=it)$ (black line) and $\mathrm{Re}f'(z=it)$ (red line) for the quench $b=4.1\to 3.9999$ for a system with $N=6001$ $\big(\Delta k = \frac{2\pi}{Na}\big)$ a true 1D tail with a 0D critical time as shown in Fig.~\ref{swsmzero}(f), showing a logarithmic singularity at the 0D critical time.}
\label{swsmfreeE}
\end{center}
\end{figure}

Following an analysis of the conditions as well as the critical times for the occurrence of DQPTs, it remains to study the nature of the non-analyticities observed in the dynamical free energy. Further, recent achievements in the experimental observation of dynamical phase transitions~\cite{PhysRevLett.119.080501}, in particular the dynamical free energy, renders its study a promising way to analyze the zeros of the boundary partition function and the critical times in an experimental setup. The dimension of the zeros of the boundary partition function, as seen in the previous section, remarkably manifest themselves in the nature of the non-analyticities of the dynamical free energy. In Fig.~\ref{swsmfreeE} we show the free energy as given in~\eqref{dqptf}, integrated over the 2D Brillouin zone at $k_z=0$. We only show it for the quench described in Fig.~\ref{swsmzero}(a) as it includes all the relevant features. It is found that there are indeed two distinct kinds of zero structures namely, the 1D and the 2D tails, which result in critical times of different dimensions arising from the intersections of the tails with the imaginary axis. To ascertain the dimension conclusively we resort to an analysis of the dynamical free-energy. A 1D zero structure intersecting with the imaginary axis should result in discontinuities in the first derivative of the dynamical free energy, while the 2D zero structures with 1D imaginary axis intersections should lead to non-differentiable points with discontinuities in the second derivative of the dynamical free energy~\cite{PhysRevB.92.075114}. From Fig.~\ref{swsmfreeE}(a), we see that $\mathrm{Re}f'(t)$ is non-differentiable at the edges of the shaded regions, which depict the 1D imaginary axis intersections of the 2D zero manifolds. Further, in Fig.~\ref{swsmfreeE}(b) we zoom into the region near the first seemingly 0D critical time shown by the green dashed line in Fig.~\ref{swsmfreeE}(a), revealing that $\mathrm{Re}f'(t)$ has two points of non-differentiability. Hence, it is concluded that the thin tails are in fact 2D structures with 1D critical times arising from the imaginary axis intersections. 

Now, in Fig~\ref{swsmfreeE}(c), for a quench beginning at $b=4.1$ in the $C=0$ phase to the $C=1$ phase with the quench distance governed by~\eqref{deltab0d}, a singular behavior is observed in $\mathrm{Re}f(t)$, concurrent with the discussion in the previous section where it was shown that the 0D critical time is isolated. Unlike a dense 1D tail of zeros which leads to a non-analyticity in $\mathrm{Re}f(t)$ of the form $\mathrm{Re}f(t)=\frac{|t-t_c|}{\tau}$, an isolated zero leads to a logarithmic singularity of the form $\mathrm{Re}f(t)\sim \mathrm{log}(|t-t_{0D}|)$, where $t_{0D}$ is the isolated 0D critical time. 

Therefore, with suitably implemented quench protocols, both 1D and 0D critical times may be observed in a true 2D system (restriction to a given value of $k_z$ leads to an effective 2D system), with the free-energy density serving as an indicator to probe the nature of the zeros of the dynamical partition function. This is in contrast to 0D critical times obtained in the Kitaev honeycomb model~\cite{PhysRevB.92.075114} by an explicit reduction of the system dimension. 

\section{Dynamical vortices}
\label{secdynvort}
Time- and momentum-resolved full state tomography permits the extensive mapping of the entire wavefunction of a non-interacting ultracold fermionic gas in driven optical lattices. Recent studies have experimentally~\cite{Flaschner2018} and theoretically~\cite{PhysRevA.96.023601} demonstrated the appearance, movement and annihilation of dynamical azimuthal Bloch phase vortices in momentum space following sudden quenches across a topological phase boundary using state tomography. These dynamical vortices have been interpreted as the dynamical Fisher zeros of the LO, holding immense significance as they provide a way of mapping the dynamical Fisher zeros in real time. Other schemes have also attempted the same, in particular the one by Brandner \textit{et al.}~\cite{PhysRevLett.118.180601} obtain the zeros of stochastic Andreev tunneling between a normal-state island and two superconducting leads from measurements of the dynamical activity along a trajectory. We explain the conditions for the appearance and dynamics of such vortices in detail with an emphasis on WSMs, while also deriving the role of the dimension of the critical times or equivalently the dimension of the imaginary axis intersections of the zeros of the boundary partition function. It is discovered that the dynamical vortices are not a universal and infallible indicator of the zeros of the boundary partition function as only 2D zeros with 1D critical times are manifested as dynamical vortices. 

The time-evolved state of a two-band Hamiltonian may be visualized on the relative Bloch sphere for each quasi-momentum, where the south pole corresponds to the ground state of the initial Hamiltonian. Beginning with the ground state of the initial Hamiltonian, the time-evolved state following a sudden quench is given by,
\begin{align}
\lvert g_i\rangle &\xrightarrow{\mathrm{quench}}u|g_f\rangle+v|e_f\rangle\xrightarrow{t>0}ue^{ih_ft}|g_f\rangle+ve^{-ih_ft}|e_f\rangle\nonumber\\
&=\left(u^2e^{ih_ft}+v^2e^{-ih_ft}\right)|g_i\rangle+uv\left(e^{ih_ft}-e^{-ih_ft}\right)|e_i\rangle\label{relblochitevol}
\end{align}
where the subscripts $i(f)$ refer to the pre-quench(post-quench) systems and the obvious $\bm{k}$ dependence of all quantities has been suppressed for the sake of brevity. For any given wavevector, a DQPT occurs when the time-evolved state at the critical time is orthogonal to the post-quench state at $t=0$, or equivalently when the time-evolved state reaches the north pole of the relative Bloch sphere. This may only occur if $\bm{h_f}\cdot\bm{h_i}=0$. Note that the time evolution by the post-quench Hamiltonian is represented in the relative Bloch sphere by a precession about the post-quench Hamiltonian vector, with the initial state oriented along the south pole. For $\bm{h_f}\cdot\bm{h_i}=0$, a state with critical $\bm{k}^*:p_{\bm{k}^*}=\frac{1}{2}$ reaches the north pole at $t=t_c$ by traversing a longitude or equivalently the circumference of the Bloch disc normal to $\bm{h_f}(\bm{k}^*)$. Since the time evolution is governed by unitary dynamics with a smooth spectrum ($\partial_{kx}h_f$ and $\partial_{ky}h_f$ exist and are continuous), for $\bm{k}\neq\bm{k}^*$ lying in $\mathcal{C}=\{\bm{k}:|\bm{k}-\bm{k_c}|<\epsilon\lvert_{\epsilon\to 0^+}\}$ we have $\bm{h_f}(\bm{k})=\bm{h_f}(\bm{k}^*)+\nabla\bm{h_f}(\bm{k}^*)\cdot\delta\bm{k}=\bm{h_f}(\bm{k}^*)+\mathcal{O}(\delta k)$. Therefore, the neighboring $\bm{k}$ states reach near the north pole by traversing the circumferences of slightly tilted Bloch discs normal to $\bm{h_f}(\bm{k})$ with the polar-angular separation nearly equaling $\mathrm{cos}^{-1}\left(\frac{\bm{h_f}(\bm{k})\cdot\bm{h_f}(\bm{k}^*)}{h_f(\bm{k})h_i(\bm{k})}\right)$ (since the spectrum is smooth but not constant, minor variations $\sim\mathcal{O}(\delta k)$ occur as neighboring $\bm{k}$ states haven't yet traversed half the circumference of their Bloch discs). This guarantees the existence of this patch. 

The presence of a phase vortex is inferred through the vorticity of the phase field, given by $\oint d\bm{k}\cdot\nabla\phi=\iint d^2\bm{k}\nabla\times\nabla\phi$. In general, the curl of the gradient of a scalar function in simply-connected domains is zero in the absence of singularities of the scalar function. Therefore, the phase, which is defined everywhere except at the zeros of $\bm{v}$, harbors singularities only at the zeros~\cite{Dennisthesis,Dennis,berry1,berry2} of the field $\bm{v}$. In a region of 2D-space the zeros of $\bm{v}$ are counted by,
\begin{align}
\iint dv_1dv_2 \delta(v_1)\delta(v_2)&=\iint d^2\bm{r}\left\lvert\frac{\partial(v_1,v_2)}{\partial(r_1,r_2)}\right\rvert\delta(v_1)\delta(v_2)\nonumber\\
&=\iint d^2\bm{r}\sum_{\bm{v}(\bm{r_a})=0}\delta^2(\bm{r}-\bm{r_a})  \label{nozeros}
\end{align}  
The net vorticity or topological charge of these point phase defects are given by weighting the number of zeros in~\eqref{nozeros} with the corresponding topological charge, which is the winding number of
the complex field $\bm{v}$ for an infinitesimal closed curve around the defect. We find the vorticity of a 2D vector field $\bm{v}=v_1+iv_2=ve^{i\phi}$ parameterized by $(k_x,k_y)$, for which $\nabla\phi = \frac{v_1\nabla v_2-v_2\nabla v_1}{v^2}=\epsilon_{ab}n^a\nabla n^b$ where $n^{a(b)}=\frac{v_{a(b)}}{v}$. Here $\nabla\phi$ is similar to the superfluid velocity. Now, on expanding $\bm{v}$ at $k^*$, we get $\bm{v}(\bm{k})=ve^{i\phi}=\bm{v}(\bm{k^*})+\nabla\bm{v}(\bm{k^*})\cdot\bm{\delta k}=\eta_1+i\eta_2 + \zeta_1+i\zeta_2$, where $\bm{v}(\bm{k^*})=\eta_1+i\eta_2$, $\nabla\bm{v}(\bm{k^*})\cdot\bm{\delta k}=\zeta_1+i\zeta_2$ and $v=\sqrt{(\eta_1+\zeta_1)^2+(\eta_2+\zeta_2)^2}$. The vorticity at any $\bm{k}$ is found by integrating in an infinitesimal loop around $\bm{k_c}$ as given by $u_s=\oint \bm{v}\cdot d\bm{k}=\iint \nabla\times\bm{v}d^2\bm{k}$. Since the path chosen is irrelevant in a curl-free region not enclosing $\bm{k_c}$, one may choose a circle $\bm{k}=\bm{k_c}+\delta k(\mathrm{cos}(\alpha), \mathrm{sin}(\alpha))$, where $\alpha\in[0,2\pi)$ and $\delta k$ is infinitesimal. Hence, 
\begin{align}
\partial_\alpha\phi&=\frac{v_1\partial_\alpha v_2-v_2\partial_\alpha v_1}{v^2}=\frac{\zeta_1\partial_\alpha\zeta_2-\zeta_2\partial_\alpha\zeta_1}{(\eta_1+\zeta_1)^2+(\eta_2+\zeta_2)^2}\nonumber\\
&+\frac{\eta_1\partial_\alpha\zeta_2-\eta_2\partial_\alpha\zeta_1}{(\eta_1+\zeta_1)^2+(\eta_2+\zeta_2)^2}-i\frac{\eta_1\partial_\alpha\zeta_1+\eta_2\partial_\alpha\zeta_2}{(\eta_1+\zeta_1)^2+(\eta_2+\zeta_2)^2}\nonumber\\
u_s&=\frac{1}{2\pi}\displaystyle{\lim_{\delta k\to 0}}\int_{0}^{2\pi} d\alpha\partial_\alpha\phi \label{vortgen}
\end{align}
Note that, $\eta_{1,2}\sim\mathcal{O}(1)$ while $\zeta_{1,2}\sim\mathcal{O}(\delta k^2)$. Hence, in the limit $\delta k\to 0$, the vorticity ($u_s$) vanishes unless $\eta_1=\eta_2=0$, i.e. $\bm{v}(\bm{k^*})=0$. This is expected as $\nabla\times\nabla\phi=0$ in regions devoid of singularities of $\nabla\phi$ or equivalently zeros of $\bm{v}$. On expanding about a zero of $\bm{v}$, 
\begin{align}
&\bm{v}(\bm{k})=\underbrace{\bm{v}(\bm{k^*})}_{=0}+\nabla\bm{v}(\bm{k})\cdot\bm{\delta k}\nonumber\\
&=\left(\frac{\partial v_1}{\partial k_x}\mathrm{cos}(\alpha)+\frac{\partial v_1}{\partial k_y}\mathrm{sin}(\alpha),\frac{\partial v_2}{\partial k_x}\mathrm{cos}(\alpha)+\frac{\partial v_2}{\partial k_y}\mathrm{sin}(\alpha)\right)\nonumber
\end{align}
On substituting this in~\eqref{vortgen} we get,
\begin{align}
u_s&=\frac{1}{2\pi}\frac{\partial(v_1,v_2)}{\partial(k_x,k_y)}\int_{0}^{2\pi} \frac{d\alpha}{\left(\splitdfrac{\left(\frac{\partial v_1}{\partial k_x}\mathrm{cos}(\alpha)+\frac{\partial v_1}{\partial k_y}\mathrm{sin}(\alpha)\right)^2}{+\left(\frac{\partial v_2}{\partial k_x}\mathrm{cos}(\alpha)+\frac{\partial v_2}{\partial k_y}\mathrm{sin}(\alpha)\right)^2}\right)}\nonumber\\
&=\hspace*{0.1cm}\mathrm{sgn}\frac{\partial(v_1,v_2)}{\partial(k_x,k_y)} \label{vortgenres}
\end{align} 

\begin{figure*}[htb!]
\begin{center}
\includegraphics[width=.84\linewidth]{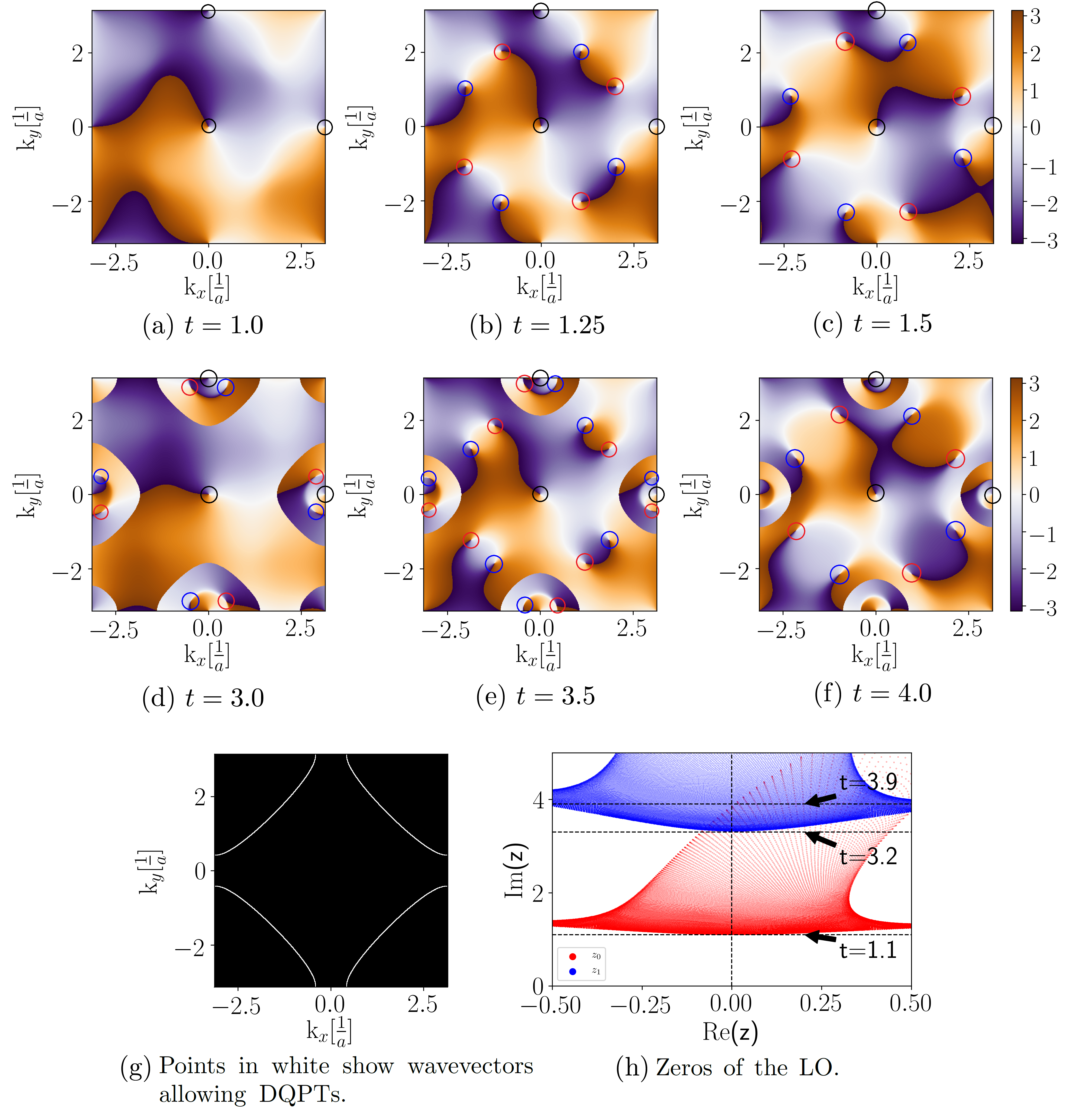}
\caption{(a)-(f) Azimuthal phase map for the quench described by $t=-t'=1$, $k_z=0$ and $b=\infty\to1.9$. Static vortices are marked in black while the dynamical vortices with positive(negative) vorticity are marked in red(blue). The static vortices appearing at $(k_x,k_y)=(0,\pm\pi)$ and $(\pm\pi,0)$ continue over the extended BZ and thus represent the same static vortices. The vortices develop and move along the $p_{\bm{k}}=\frac{1}{2}$ contours in the 2D BZ, appearing initially in the middle of each $p_{\bm{k}}=\frac{1}{2}$ contour and moving outwards towards the edge of the BZ with increasing time. (g) $p_{\bm{k}}=\frac{1}{2}$ contours shown in white. (h) The zeros $z_0$ and $z_1$, with the boundaries of the corresponding critical times marked.}
\label{dqptvort}
\end{center}
\end{figure*}

where $\frac{\partial(v_1,v_2)}{\partial(k_x,k_y)}$ is the Jacobian of the transformation $(k_x,k_y)\to(v_1,v_2)$. This implies that,$\nabla\times\nabla\phi= 2\pi\delta^2(\bm{v})(\nabla v_1\times\nabla v_2)=\sum_{j}\delta(\bm{k}-\bm{k_j})\hspace*{0.1cm}\mathrm{sgn}(\nabla v_1\times\nabla v_2)=\sum_{j}\delta(\bm{k}-\bm{k_j})\hspace*{0.1cm}\mathrm{sgn}\frac{\partial(v_1,v_2)}{\partial(k_x,k_y)}$ where $\mathrm{sgn}\frac{\partial(v_1,v_2)}{\partial(k_x,k_y)}$ is the topological charge of the vortex and $\bm{k_j}$ are the zeros of $\bm{v}$. Note that the system may also admit lines of zeros of $\bm{v}$ instead of isolated points if the contours of $v_1$ and $v_2$ are matched or equivalently $\nabla v_1\propto \nabla v_2$ at the zeros of $\bm{v}$. This implies that the Jacobian $\frac{\partial(v_1,v_2)}{\partial(k_x,k_y)}$ vanishes and therefore $\nabla\times\nabla\phi=0$, resulting in an absence of vortices.

Now, we analyze the time-evolved post-quench state. First, considering pre-quench $b\to\infty$ for which $\lvert g_i\rangle=[0,1]^T$, we get, 
\begin{align}
\lvert\Psi_f\rangle&=\mathrm{cos}(h_f t)\lvert g_i\rangle-i\mathrm{sin}(h_f)\bm{\sigma}\cdot\hat{\bm{h_f}}\lvert g_i\rangle\nonumber\\
&=\begin{bmatrix}i\mathrm{sin}(h_ft)(\hat{h_x}-i\hat{h_y})\\\mathrm{cos}(h_ft)+i\mathrm{sin}(h_ft)\hat{h}_{f,z}\end{bmatrix} \label{tevolpreinfi}
\end{align}
Now, casting this into the Bloch state form $[\mathrm{sin}\left(\frac{\theta}{2}\right),-\mathrm{cos}\left(\frac{\theta}{2}\right)e^{i\phi}]$, the azimuthal phase is obtained as $\phi_d-\phi_n$ where $\phi_{d(n)}$ is the phase of the bottom(top) element of the time evolved state spinor $\lvert\Psi_f\rangle$. The vector field in the upper element $\hat{h_y}(\bm{k^*})+i\hat{h_x}(\bm{k^*})\neq 0$ and thus its vorticity vanishes. However, the vector field in the lower element of~\eqref{tevolpreinfi} admits phase vortices as it vanishes at $\bm{k^*}={\bm{k^*}:\hat{h_z}=0}$ and $t=t^*=(2n+1)\frac{\pi}{2h_f(k^*)}$ at which $\langle g_i|\Psi_f(\bm{k^*},t^*)\rangle=\frac{1}{2}$. This is exactly the condition for the occurrence of DQPTs, thus establishing the correspondence between the appearance of the dynamical vortices and the zeros of the LO. Using~\eqref{tevolpreinfi},~\eqref{vortgen} and~\eqref{vortgenres} along with $v_1=h_f$ and $v_1=\hat{h_{f,z}}$,
\begin{align}
u_s^{\mathrm{dynamic}}&=\hspace*{0.1cm}\mathrm{sgn}\left(\frac{\partial(h_f,\hat{h}_{f,z})}{\partial(k_x,k_y)}\right) \label{dynvort}
\end{align}
which is evaluated at $\bm{k}=\bm{k^*}$. Now, at $\bm{k'}\neq\bm{k^*}$ when $h_x(\bm{k'})=h_y(\bm{k'})=0$, the contribution of the azimuthal phase of the upper component $\phi_u$ to the net azimuthal phase $\phi=\phi_u+\phi_l$ results in vortices, while the lower component yields no vortices. These static vortices, which have no connection with the occurrence of DQPTs, remain stationary with time. The appearance of static vortices are contingent upon the existence of wavevectors $\bm{k'}$ for which $h_x(\bm{k'})=h_y(\bm{k'})=0$, which in our model of WSM are $(k_x,k_y)=(0,0), (\pm\pi,0), (0,\pm\pi), (\pm\pi,\pm\pi)$. The time-evolved states with wavevectors close to $\bm{k'}$ occupy a patch enclosing the south pole, resulting in an azimuthal phase vortex pinned at $\bm{k'}$. An analysis similar to the one carried out for dynamical vortices yields the vorticity for the static vortices, 
\begin{align}
u_s^{\mathrm{static}}&=-\hspace*{0.1cm}\mathrm{sgn}\left(\frac{\partial(\hat{h}_{f,y},\hat{h}_{f,x})}{\partial(k_x,k_y)}\right)=\hspace*{0.1cm}\mathrm{sgn}\left(\frac{\partial(\hat{h}_{f,x},\hat{h}_{f,y})}{\partial(k_x,k_y)}\right) \label{statvort}
\end{align}
Therefore, the net vorticity is given by $u_s=u_s^{\mathrm{static}}+u_s^{\mathrm{dynamic}}=\mathrm{sgn}\left(\frac{\partial(h_{f},\hat{h}_{f,z})}{\partial(k_x,k_y)}\right)+\mathrm{sgn}\left(\frac{\partial(\hat{h}_{f,x},\hat{h}_{f,y})}{\partial(k_x,k_y)}\right)$.

The azimuthal phases of the time-evolved state vectors over the entire 2D BZ is shown in Fig.~\ref{dqptvort}. The dynamical vortices are found to exhibit two characteristic features. First, as expected, the dynamical vortices in Fig.~\ref{dqptvort} (a)-(i) exist and move on the $p_{\bm{k}}=\frac{1}{2}$ lines shown in Fig.~\ref{dqptvort}(h). The motion of the dynamical vortices depends on the energy ($h_f$) of the participating states with $p_{\bm{k}}=\frac{1}{2}$. Vortices appear first at the state with the largest $h_f$ and move along the $p_{\bm{k}}=\frac{1}{2}$ line towards smaller values of $h_f$ with vortices of opposite vorticity moving in opposite directions. From the energy dispersion of the WSM model, it is clear that the maximas of $h_f$ along the $p_{\bm{k}}=\frac{1}{2}$ lines shown in Fig.~\ref{dqptvort}(h) occur where the lines intersect $|k_y|=k_x$, while the minimas occur at $k_x=0$ or $k_y=0$. Hence, the vortices emerge at the center of the $p_{\bm{k}}=\frac{1}{2}$ lines and move outwards. Further, at $t=3.5$ where the critical times from both sets of zeros $z_0$ and $z_1$ are active, two sets of dynamical vortices are observed corresponding to each set of zeros. The dynamical vortices near the center of the $p_{\bm{k}}=\frac{1}{2}$ curve correspond to $z_1$ which has just started while the vortices nearing the BZ edges correspond to $z_0$ which is nearing its end. Second, we find that the dynamical vortices are not a faithful manifestation of all zeros of the LO. As noted earlier, for the $b_1\to\infty$ case the wave-vectors permitting DQPTs are the solutions of $\hat{h}_{f,z}=0$. When the contours of $\hat{h}_{f,z}$ are aligned with the contours of $h_f$, we get a loop of wave-vectors permitting DQPTs in the 2D Brillouin zone with the same value of $h_f$. Note that this leads to 0D critical times as all the states eligible to yield a DQPT do so simultaneously ($t_c\propto\frac{1}{h_f(\bm{k}^*)}$). Now, since the contours of $v_1=h_f$ and $v_2=\hat{h}_{f,z}$ are aligned, $u_s^{\mathrm{dynamic}}\propto\mathrm{sgn}\frac{\partial(v_1,v_2)}{\partial(k_x,k_y)}=\mathrm{sgn}(\nabla v_1\times\nabla v_2)=0$. Hence, 0D critical times formed by imaginary axis intersections of 1D zeros of the boundary partition function can not lead to phase vortices. This includes quenches from the upper $C=0$ phase to the $C=1$ phase for which the zeros of the LO lie around $(k_x,k_y)=(0,0)$ and overlap with the contours of $h_f$. Note that, the dynamical free energy density does permit an identification of the 0D critical times as shown in Fig.~\ref{swsmfreeE} and the corresponding discussion. Clearly, a combination of these two methods, namely, dynamical phase vortices and dynamical free energy density, provide an experimentally realizable way to probe the structure and dimension of the critical times. The conclusions of this argument are independent of the quench parameters as long as the condition of 0D critical time is met. 

\section{Conclusion}
In this work we have explored quenches in type-I inversion symmetric Weyl semimetals and the consequent emergence of dynamical phase transitions. We find conditions for the occurrence of dynamical quantum phase transitions (DQPT), connecting them with the equilibrium topological properties of the system namely, the Chern number characterizing the ground state. We discover that a change in the signed Chern number is a sufficient but not necessary condition for the occurrence of DQPTs by analyzing the ground state fidelity. Additionally, by tuning the ratio of the transverse and longitudinal hopping parameters, DQPTs may also be triggered for quenches lying entirely within the initial Chern phase. Moreover, we analyze the zeros of the boundary partition function discovering that in general, the zeros form two-dimensional structures resulting in one-dimensional critical times. However, quenches near the Chern phase boundaries appropriately scaled with the system size may lead to one-dimensional structures with zero-dimensional critical times. These are manifested as logarithmic singularities in the dynamical free energy. Finally, following recent developments in Bloch state tomography through which dynamical Fisher zeros of the Loschmidt overlap (critical times for the occurrence of DQPTs) are observed through azimuthal Bloch phase vortices, we investigate the same in Weyl semimetals. We establish the role of the dimension of the critical times while deriving the vorticity, discovering that only one-dimensional manifolds of critical times create dynamical vortices. Lastly, the results of this work may be conveniently extended to Chern insulators due to the similarity of the corresponding Hamiltonians.

\indent {\it{Acknowledgments:}} SB thanks Markus Heyl and
Daniele Trapin for many discussions and acknowledges support from DST,
India, through Ramanujan Fellowship Grant No. SB/S2/RJN-128/2016.
\bibliographystyle{apsrev}
\bibliography{bibl}
\end{document}